\documentclass[fleqn,10pt]{wlscirep}
\usepackage[utf8]{inputenc}
\usepackage[T1]{fontenc}
\usepackage{lineno}
\usepackage{cite}
\usepackage{amsmath,amssymb,amsfonts}
\usepackage{graphicx,booktabs,array,chemformula}
\usepackage{algorithmic}
\usepackage{graphicx}
\usepackage{textcomp}
\usepackage{placeins}
\usepackage{color}
\usepackage{subfig}
\usepackage{acronym}
\usepackage{appendix}
\usepackage{float}
\usepackage{adjustbox}
\linethickness{0.5cm}

\usepackage{lineno}
\definecolor{red}{rgb}{1.00, 0.00, 0.00}
\definecolor{green}{rgb}{0.00, 1.00, 0.00}
\definecolor{yellow}{rgb}{1.00, 0.67, 0.00}
\definecolor{blue}{rgb}{0.00, 0.00, 1.00}
\definecolor{violet}{rgb}{0.67, 0.67, 1.00}
\usepackage{pifont}
% ...
% ...
\title{TRUSTED: The Paired 3D Transabdominal Ultrasound and CT Human Data for Kidney Segmentation and Registration Research}

\author[1,2, *]{William Ndzimbong}
\author[3]{Cyril Fourniol}
\author[4]{Loic Themyr}
\author[5]{Nicolas Thome}
\author[2]{Yvonne Keeza}
\author[6]{Beno\^it Sauer}
\author[7]{Pierre-Thierry Piéchaud}
\author[8]{Arnaud Méjean}
\author[9]{Jacques Marescaux}
\author[1]{Daniel George}
\author[10]{Didier Mutter}
\author[2,9, $\dag$]{Alexandre Hostettler}
\author[2,9, $\dag$, *]{Toby Collins}

\affil[1]{University of Strasbourg, CNRS, ICUBE Laboratory, Strasbourg, France}
\affil[2]{Research Institute against Digestive Cancer (IRCAD), Kigali, Rwanda}
\affil[3]{Department of Urology, HEGP, APHP, Paris, France}
\affil[4]{Conservatoire National des Arts et Métiers (CNAM), CEDRIC, Paris, France}
\affil[5]{Sorbonne University, CNRS, ISIR, Paris, France}
\affil[6]{Department of Radiology, Clinique Sainte-Anne, Strasbourg, France}
\affil[7]{Clinique Saint Augustin, ELSAN, Bordeaux, France}
\affil[8]{H\^opital Europ\'een Georges-Pompidou, Paris, France}
\affil[9]{Research Institute against Digestive Cancer (IRCAD), Strasbourg, France}
\affil[10]{Institute of Image-Guided Surgery (IHU), Strasbourg, France}

\affil[*]{corresponding authors: William Ndzimbong (william.ndzimbong@ircad.fr), Toby Collins (toby.collins@ircad.fr)}
\affil[$\dag$]{these authors contributed equally to this work}

\begin{abstract}
Inter-modal image registration (IMIR) and image segmentation with abdominal Ultrasound (US) data has many important clinical applications, including image-guided surgery, automatic organ measurement and robotic navigation. However, research is severely limited by the lack of public datasets.
We propose TRUSTED (the Tridimensional Renal Ultra Sound TomodEnsitometrie Dataset), comprising paired transabdominal 3DUS and CT kidney images from 48 human patients (96 kidneys), including segmentation, and anatomical landmark annotations by two experienced radiographers. Inter-rater segmentation agreement was over 94 (Dice score), and gold-standard segmentations were generated using the STAPLE algorithm. Seven anatomical landmarks were annotated, important for IMIR systems development and evaluation. To validate the dataset's utility, 5 competitive Deep Learning models for automatic kidney segmentation were benchmarked, yielding average DICE scores from 83.2\% to 89.1\% for CT, and 61.9\% to 79.4\% for US images. Three IMIR methods were benchmarked, and Coherent Point Drift performed best with an average Target Registration Error of 4.53mm. The TRUSTED dataset may be used freely by researchers to develop and validatate new segmentation and IMIR methods.
\end{abstract}
\begin{document}

\flushbottom
\maketitle

\thispagestyle{empty}

% \noindent Please note: Abbreviations should be introduced at the first mention in the main text – no abbreviations lists or tables should be included. Structure of the main text is provided below.

\section*{Background \& Summary}
%\textcolor{red}{(700 words max)}

% (700 words maximum) An overview of the study design, the assay(s) performed, and the created data, including any background information needed to put this study in the context of previous work and the literature. The section should also briefly outline the broader goals that motivated the creation of this dataset and the potential reuse value. We also encourage authors to include a figure that provides a schematic overview of the study and assay(s) design. The Background \& Summary should not include subheadings. This section and the other main body sections of the manuscript should include citations to the literature as needed. 

Ultrasound (US) is used in many clinical applications, especially image-guided surgery, disease diagnosis, and obstetrics. Compared to Computed Tomography (CT) and Magnetic Resonance (MR), its various advantages include higher portability, lower cost, real-time imaging and no ionizing radiation. However, US is more operator dependence, has higher image noise, and image artifacts \cite{european2019abdominal}. Inter-modal image registration (IMIR) can help to overcome those limitations, by spatially aligning (registering) US with CT or MR images. The aligned images can allow operators to locate targets intra-operatively that may be difficult to see using only US, such as renal lesions. 

IMIR is not a solved problem, and proposed algorithms mainly fall into two categories: surface-based and intensity-based. Surface-based methods align the organ surfaces, extracted from the images using a segmentation method (see below). Early approaches used manual segmentation \cite{leroy2006percutaneous}, which has limited practical utility. More recently, automatic image segmentation (i.e. structure delineation in medical images)  has been used \cite{fu2021biomechanically}. Registration may be achieved with various algorithms, including Iterative Closest Point (ICP) \cite{icp1987}, GMMTree \cite{Agamennoni2016}, Bayesian Coherent Point Drift (BCPD) \cite{hirose2021}, or DNNs such as PointNetLK \cite{yaoki2019} and PCRNet \cite{Sarode2019}. In contrast, intensity-based IMIR methods align the images directly using voxel intensity information. Common approaches iteratively optimize a cost function such as Local Normalized Cross Correlation (LNCC) \cite{Baig2012} and Local Cross-Correlation (LC2)  \cite{den1993calculation, hale2006fast}, or a cost function learned using a DNN \cite{haskins2019learning}. DNNs have been developed to directly register medical images, such as Quicksilver \cite{Yang2017}, LocalNet \cite{Hu2018, hu2018weakly}, VoxelMorph \cite{voxelmorph2019}, TransMorph \cite{CHEN2022102615}. The challenges for DNN-based approaches include assembling sufficient paired inter-modal training data, domain shifts, and handling image quality variability (especially important with US data). 

Automatic US image segmentation has other key clinical applications in additon to IMIR \cite{chen2021mr, fu2021biomechanically, leroy2006percutaneous, karnik2010assessment}, including structure volume measurement, improved visualization, disease monitoring, such as chronic kidney disease \cite{shruthi2015detection}, and in robotic systems for delicate structures avoidance \cite{kettenbach2005robot}. DNNs achieve best image segmentation performance, popular models being nnUNet \cite{Isensee2020}, TransUnet \cite{chen2021}, Swin-UNet \cite{cao2023}, nnFormer \cite{zhou2021}, CoTr \cite{cotr2021}, Glam \cite{themyr2023}. These models are regularly evaluated for segmentation in human abdominal CT images with public datasets \cite{Yu2019, Gibson2018, zhou2018, Isensee2020, chen2021, cao2023, zhou2021, cotr2021, themyr2023}, and CT segmentation challenges at \textit{e.g.} MICCAI 2020, MICCAI 2021 and AMOS 2022. However, evaluation on abdominal US data have been exclusively based on private datasets \cite{boussaid2021shape, zeng2020label, xu2022polar}, restricting results reproduction. Indeed, research in both IMIR and segmentation, with abdominal US data, has been highly limited by the total lack of annotated public datasets.  

We introduce the TRUSTED Dataset; the first public dataset of paired transabdominal 3DUS and CT images of human patients with manually segmented organs. The dataset has 96 kidneys from 48 patients, and two experienced radiologists (10 and 11 years) annotated each image with kidney segmentation, and locations of seven anatomical landmarks per kidney. Gold-standard segmentation was computed by fusing the segmentations with the STAPLE algorithm \textcolor{red}{REF}. Most related datasets provide only one annotation per image, Here, two independent annotations were used to quantify annotation agreement for quality reporting, and also for future research to in investigate how annotation agreement relates to algorithm output uncertainty.

Combined with standard data augmentation techniques, the TRUSTED dataset is of a sufficient size to train and evaluate DNN models. To demonstrate its research utility, five DNN segmentation methods are benchmarked for automatic CT and US kidney segmentation: 3DUNe \cite{Cicek2016}, VNet \cite{Milletari2016}, nnUNet \cite{Isensee2020}, CoTr \cite{cotr2021} and Glam \cite{themyr2023}, demonstrating a range of performances particularly on US data. We also benchmarked two surface-based IMIR methods (using automatic kidney surface segmentation): ICP \cite{icp1987} and BCPD \cite{hirose2021}, and two intensity-based methods \cite{horstmann2022orientation, markova2022global}, each using rigid and affine transform models settings. Additionally, we show the dataset can be used to conduct a registration initialization sensitivity analysis, which is important to fully characterize the performance of IMIR methods. 

The purpose of these demonstrations is to show the usefulness of the dataset for research purposes. Certainly, additional methods can be compared by other researchers, especially to validate new segmentation and IMIR algorithms when they are submitted for publication in technical conferences or journals, such as MICCAI, IPCAI, and TMI. Our closest works are the MICCAI 2023 Mu-RegPro challenge \cite{Baum2023}, and Fedorov \textit{et al.} \cite{Fedorov2015}, both containing trans-rectal 3DUS and MR images of the prostate, but not the kidney. Importantly, not only do they involve a different organ, they also do not represent the difficulties of segmenting and registering transabdominal US data, especially obstructions due to image artefacts and low contrast due to visceral fat, and acoustic shadows from ribs and intestines. %As a consequence, a dedicated public dataset, proposed here, for segmentation and IMIR with transabdominal US data is important, which directly motivates this work.

\section*{Methods}

\subsection*{Data Collection}
The 3DUS and CT images of this database were collected prospectively in a single-center study called ``Tridimentional Renal Ultra Sound TomodEnsitometrie Database" (TRUSTED) (National ID RCB: 2020-A01029-30/SI:20.05.01.539110).  This was approved by the French government's "Comité de Protection des Personnes Sud-Mediterranée II". The study enrolled 48 men and women aged between 18 and 80 who were scheduled to undergo a standard abdominal or abdominal-pelvic CT scan as part of their routine care at the Nouvel Hôpital Civil (NHC) of Strasbourg, France. Each participant enrolled in the study with written informed consent.
CT images were acquired in three phases (no-contrast, venus, and portal), with dimensions $512 \times 512 \times (586\pm 219)$ voxels and $(0.81\pm 0.08)$ mm $\times~(0.81\pm 0.08)$ mm $\times~(1.05\pm 0.28)$ mm voxel spacing. Immediately after the CT scan (Siemens Somatom Force 384), the patient underwent a transabdominal US examination using two US probes: a 2D convex transabdominal probe (reg), and a 3D transabdominal probe (Seimens Acuson S3000 7CF1 HD, mechanically driven). The 3D probe acquired 3DUS images in approximately 2 seconds with a resolution of $760 \times 540 \times (739\pm 55.15)$ voxels with an isotropic voxel spacing of $0.3$ mm. The patient's left and right kidneys were imaged with the transabdominal 3DUS probe, with $4 \pm 2$ image acquisitions per patient. All images were de-identified by removing sensitive text from image contents and DICOM metadata. 

\subsection*{Data Annotation}
\label{sec:data_annotation}
\begin{figure}[!h]
    \begin{center}
        \begin{tabular}{c@{\hspace{1pt}}c@{\hspace{1pt}}c}
            & Manual segmentation & Manual landmarks \\
            \rotatebox{90}{\hspace{13pt} CT, KDY\_01} & \includegraphics[height=0.17\textwidth, width=0.22\textheight]{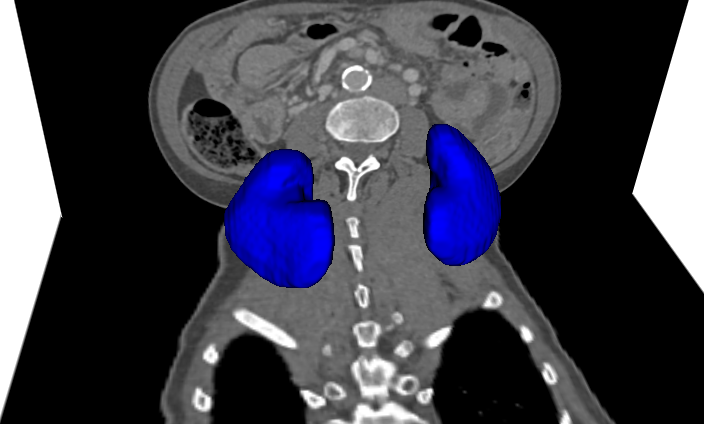} &
            \includegraphics[height=0.17\textwidth, width=0.10\textheight]{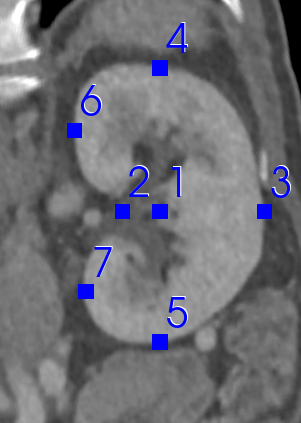} \\
            \rotatebox{90}{\hspace{10pt} US, KDY\_01R} & \includegraphics[height=0.17\textwidth, width=0.22\textheight]{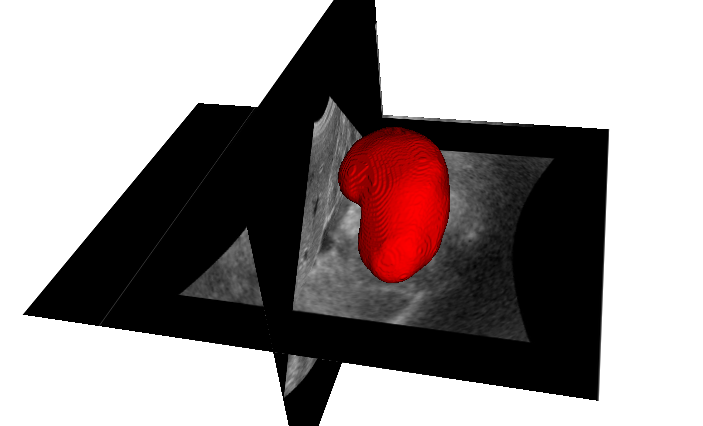} &
            \includegraphics[height=0.17\textwidth, width=0.10\textheight]{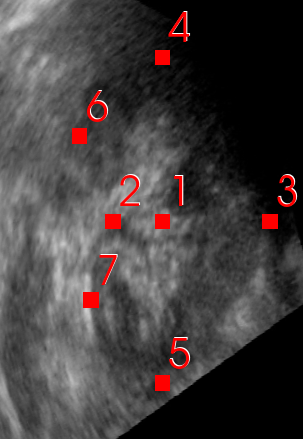} \\
        \end{tabular}
    \caption{\footnotesize Visualization of annotated data in the TRUSTED dataset (Patient 01, Annotator 2). Top left: Segmentation of left and right kidneys in CT. Top right: 2DCT slice of the right kidney in the canonical sagittal plane with 7 anatomical landmarks overlaid. Bottom left: Segmentation of the right kidney in 3DUS. Bottom right: 2DUS slice of the right kidney in the canonical sagittal plane with 7 anatomical landmarks overlaid.}
    \label{fig:manualann}
    \end{center}
\end{figure}

Each 3DUS kidney image was first inspected by an experienced radiographer (16 years). For each patient, the best left and right volumes where the kidney was fully contained within the 3D field-of-view were kept for annotation. At this stage, 59 3DUS images corresponding to the left and/or right kidneys of 39 patients were selected. All the others were excluded because their kidneys were out of the field of view. Note that, to train and evaluate DNN models, 59 3DUS images is enough to obtain state-of-the-art scores, as we will see in the "Technical validation" section, thanks in particular to the use of the data augmentation technique. The 59 3DUS selected images were annotated by two independent radiographers with 8 years of experience, using 3D Slicer \cite{Fedorov2012} (Version 4.11.20210226). We therefore provide a double annotation of the complete dataset, which is an additional contribution compared with other datasets, which almost all provide a single annotation. Each annotator worked as follows. First, a spatial transform was applied to center the kidney, followed by a 3D rotation so the kidney's principal axes aligned with the spatial axes. Next, 7 anatomical landmarks were marked on the longitudinal plane (Figure \ref{fig:manualann}, "Manual landmarks" column). The first landmark corresponded to the center of the renal pelvis and was used for Target Registration (TRE) assessment. The remaining 6 landmarks were on the kidney's surface and used for landmark-based registration initialization. The kidney was manually segmented in the axial, sagittal and coronal orientations, and 3D Slicer's volume interpolation tool was used to interactively compute the 3D kidney segmentation. Primarily, the axial orientation was used, with annotations at approximately 5 mm slice intervals. The other orientations were used to improve annotation quality especially when the boundaries in the axial orientation were unclear. Segmentation around the renal hilum is known to be imprecise  \cite{yang2018automatic}, so the annotators followed a similar definition of the renal hilum boundary as defined in past\cite{daniel2021automated}. The STAPLE algorithm \cite{staple2004}, widely used in medical imaging, was used to automatically combine the kidney manual segmentations from both radiographers, to establish ground-truth segmentations. This remains possible with the STAPLE algorithm, even if there are just 2 annotations, as indicated in the original paper \cite{staple2004}. The average landmark positions from both radiographers were used to establish ground-truth landmark positions. The CT images for all 48 patients were manually annotated with a similar workflow as US annotation, with the only difference being that both kidneys were annotated in each CT image (unlike 3DUS where only one kidney is visible in each image). Figure \ref{fig:manualann} illustrates one case of manual annotation of a CT scan and a 3DUS image by an annotator. 

The STAPLE has an important hyperparameter ($\beta$) that regulates the level of spatial smoothness. For CT, we found this parameter was not sensitive (it had very little effect on the merged segmentations) so the default value of $\beta_{CT} = 2.5$ (proposed in the original paper \cite{staple2004}) was used. However, due to a higher disagreement in US segmentation between radiographers, we found $\beta$ had a stronger influence on the US output, so it required tuning. To this end, we optimized $\beta_{US}$ based on the weakly non-linear deformation of kidney tissue  \cite{Ong2009, wein2008automatic}. A grid search was performed, where for each $\beta_{US}\in\{1.0, 2.5, 3.0, 5.0, 10.0\}$, we ran STAPLE on the US data and assessed the volume agreement (\%) between the segmentation output and the CT ground truth segmentation. We then selected $\beta_{US}$ as that which produced the highest average volume agreement, which was $\beta_{US} = 2.5$. 

\section*{Data Records}
%The TRUSTED dataset is publicly available online at (\url{https://cloud.ircad.fr/s/msTLGSHbtbicinf}). It is freely available for non-commercial research purposes. The dataset is structured as follows. \textbf{Note to reviewers} - this link is currently only available to the reviewers and the journal's editorial board, allowing them to see structure and contents of the dataset. Should this work be accepted, the link will be publicised in the final proof. Please do not distribute this link while this paper is under submission. 

The TRUSTED dataset will be made publically available pending peer review. The dataset is structured as follows:

\paragraph{Root directory:}
The root directory contains two sub-folders: 
\begin{itemize}
    \item  \textbf{CT\_DATA}, storing image and annotation data for CT images.
    \item  \textbf{US\_DATA}, storing image and annotation data for US images.
    \item  \textbf{cross\_validation\_splits.txt}, storing the patient IDs assigned to the 5-fold cross-validation folds. There are 5 rows, where row $i$ gives the patient IDs used in the test set for the $i^{th}$ fold.
    \item  \textbf{README.txt}, which provides a text description of the datatset structure.
    \item \textbf{Licence.txt}, which is a user agreement licence to be completed by anyone requesting to download this dataset. 
\end{itemize}

\paragraph{CT\_DATA directory:}
This contains four sub-directories: 
\begin{itemize}
    \item \textbf{CT\_images}, storing the CT image data in the nifti image file format \textcolor{red}{add reference}. There are 48 nifti files, which each store the CT image of one patient with \textcolor{red}{XxYxY} image resolution. Each file is an anonymized version of the raw DICOM data file produced by the CT scanner, and as such, the image intensity information unmodified with respect to the raw data.
    \item \textbf{CT\_landmarks}, storing the anatomical  landmark annotations.  This contains three sub-directories: \textbf{Annotator1} and \textbf{Annotator2} store respectively the 3D coordinates of the landmark annotations from each annotator. \textbf{GT\_estimated\_ldksCT} stores the 3D coordinates of the ground truth landmarks, computed from those of both annotators, as described in the Data Annotation section.  ground truth  In those sub-directories, there are 96 text files, each storing the 3D landmark locations of 7 anatomical landmarks for the patient's left and right kidneys. The text files are named \textit{XL1\_lktCT.txt} and \textit{XR1\_lktCT.txt}, denoting the landmark coordinates for the left and right kidneys, for patient ID X. The landmark files store the landmarks coordinates with 7 rows, and 3 columns (space delineated). The rows are sorted, so the $i^{th}$ row corresponds to coordinates of the $i^{th}$ landmark according to Fig. \ref{fig:manualann} (top right). 
    \item \textbf{CT\_masks}, storing the segmentation annotations as binary 3D masks with the same spatial resolution as the nifti image files in \textbf{CT\_images}. This contains three sub-directories: \textbf{Annotator1} and \textbf{Annotator2} store respectively the segmentation masks from each annotator. \textbf{GT\_estimated\_masksCT} stores the ground-truth segmentation masks, computed using the STAPLE algorithm, as described in the Data Annotation section. Each of these 3 sub-directories contains 48 binary nifi images, each prefixed by the patient ID. The nifti images have resolution \textcolor{red}{XxYxY}, matching exactly with the image data in \textbf{CT\_images}. In each mask file, voxel values of 0 and 255, corresponds to voxels labeled outside and inside the kidney respectively.
    \item \textbf{CT\_meshes}, storing the surface meshes, automatically extracted from the segmentation masks, using the Marching Cubes Algorithm as described in the Data Annotation section. This directory has 3 sub-directories; \textbf{Annotator1}, \textbf{Annotator2} and \textbf{GT\_estimated\_meshesCT}, which store the meshes corresponding to masks in \textbf{CT\_masks/Annotator1}, \textbf{CT\_masks/Annotator2} and \textbf{CT\_masks/GT\_estimated\_masksCT} respectively. Meshes are stored using the Waveform obj standard.     
\end{itemize}

\paragraph{US\_DATA directory:}
This directory is organized in exactly the same manner as \textbf{CT\_DATA}, and it stores all the US images and associated annotations. The same naming and file format conventions are used as with the CT data.

\section*{Technical Validation}
%\textcolor{red}{This section presents any experiments or analyses that are needed to support the technical quality of the dataset. This section may be supported by figures and tables, as needed. This is a required section; authors must present information justifying the reliability of their data.
%}

In this section, first an inter-rater agreement analysis was conducted to measure consistency between annotators. Secondly, several state-of-the-art automatic segmentation and IMIR methods are evaluated on the TRUSTED Dataset, to show its usefulness for method benchmarking, which also exposes several limitations of these methods on the clinically-relevant tasks of CT-US kidney segmentation and registration, which may be addressed by future research using this dataset.

\subsection*{Inter-Rater Agreement Analysis}
An inter-rater agreement analysis was conducted to measure consistency between annotators. This was implemented by comparing each radiographer's annotations (Ann.1 and Ann.2) against the GT annotations. Table \ref{tab:manualseg} summarises the results for segmentation, showing the mean and standard deviations (bracketed) of three evaluation metrics: Dice score, mean nearest neighbor surface distance (NN dist.), and the $95\%$ Hausdorff distance (HD95 dist) \cite{Dubuisson1994}. The results indicate a slightly higher agreement in CT segmentation compared to US, however, both CT and US segmentation agreement is generally strong. Table \ref{tab:manualland} summarises the landmark annotator agreement, showing for each landmark, the average distance between the GT and annotated positions (standard deviation in brackets). Unlike segmentation, GT landmark positions were computed as the average landmark positions from each annotator, therefore, both annotators had the same landmark distance compared to GT (as reported in Table \ref{tab:manualland}). Generally, one can see higher landmark agreement in CT, likely due to the better image contrast and, for Landmark 1 (renal pelvis), better anatomical visibility within the kidney. Concerning US, one can see a substantially lower agreement in landmarks 4-7 compared to landmarks 1-3. Recall that these landmarks were located on a saggital kidney plane, selected by the annotator. Consequently, orientation deviations of this plane led to a lever effect with higher landmark position differences at the kidney extremities.

\begin{table}[!h]
    \begin{minipage}{\linewidth}
    \centering
        \begin{tabular}{@{}ccccc@{}}
        \toprule
        Modality & Manual &  Dice score  & NN dist. & HD95 dist. \\
                 & Annotator  & in \% & in mm    & in  mm     \\
        \midrule
        CT & Ann.1 & 95.7 (1.2) & 0.61 (0.15) & 1.3 (0.43)\\ 
           & Ann.2 & 95.4 (1.1) & 0.62 (0.16) & 1.32 (0.41)\\ 
        \midrule
        US & Ann.1 & 94.4 (2.6) & 0.68 (0.41) & 1.92 (1.22)\\ 
           & Ann.2 & 94.5 (2.) & 0.67 (0.45) & 1.87 (1.37)\\ 
        \bottomrule
        \end{tabular}
    \end{minipage}
    \caption{\footnotesize Kidney segmentation inter-annotator agreement statistics for the TRUSTED dataset. For each modality (CT and US), ground-truth segmentations were computed from the annotators' segmentations (2 annotators per image, one segmentation per annotator) using STAPLE. Annotator agreement was measured by comparing the ground-truth segmentations with each annotator's segmentation (Dice, NN, and HD95 metrics). Mean and standard deviation (bracketed) are presented in the table (each annotated kidney in the dataset corresponds to one sample).}
    \label{tab:manualseg}
\end{table}

\begin{table}[!h]
    \begin{minipage}{\linewidth}
    \centering
        \begin{tabular}{@{}l@{\hspace{7pt}}l@{\hspace{7pt}}l@{\hspace{7pt}}l@{\hspace{7pt}}l@{\hspace{7pt}}l@{\hspace{7pt}}l@{\hspace{7pt}}l@{}}
        \toprule
        Modality  &  LM1 & LM2 & LM3 & LM4 & LM5 & LM6 & LM7 \\
                  &  dist. & dist. & dist. &  dist. & dist. & dist. & dist. \\
        \midrule
        CT &  1.82  &  2.84  &  2.93  &  2.10  & 2.38  & 3.55  & 2.81 \\
        &  (1.65) &  (1.81) & (2.40) & (1.27) & (1.75) & (2.64) & (1.88)\\
        \midrule
        US & 2.23  & 2.63  & 3.32  & 3.61 & 3.76 & 4.26 & 3.83\\ 
         & (1.64) & (1.98) & (2.50) & (7.20) & (7.33) & (4.34) & (4.91)\\ 
        \bottomrule
        \\
        \end{tabular}
    \end{minipage}
    \caption{\footnotesize Kidney landmark inter-annotator agreement statistics for the TRUSTED dataset. For each modality (CT and US), ground-truth 3D landmark positions were computed as the average 3D landmark position from both annotators (2 annotators per image, 7 annotated landmarks per annotator). Annotator agreement was measured by comparing the ground-truth landmark positions with each annotator's landmark position (absolute distance in mm). Mean and standard deviation (bracketed) are presented in the table. Note that by construction, the absolute distance to ground-truth was the same for each annotator.}
    \label{tab:manualland}
\end{table}

\subsection*{Automatic Segmentation}
\label{sec:comparison_seg_methods} 
\subsubsection*{Methods and Evaluation Methodology} 
Three popular CNN models have been included in this comparison: UNet \cite{Cicek2016}, VNet \cite{Milletari2016}, and nnUNet \cite{Isensee2020}, and two recent vision transformer models: CoTr \cite{cotr2021} and Glam \cite{themyr2023}.
All models were trained and evaluated using 5-fold cross-validation (same folds) while ensuring that data from the same patient was not present in different folds. To this end, the 39 patients were first randomly ordered, then they were evenly divided into each fold. The training was performed in two settings: \textit{(i)} Single Target, where the training labels corresponded to the fused manual segmentations produced by STAPLE, and \textit{(ii)} Double Target, where both manual segmentations from each annotator were used as equally-weighted training targets. The motivation to compare them was to study whether label uncertainty information, contained implicitly using Double targets, helped model performance. 

All training images were downsampled to 128-128-128 voxels to ensure all models could be trained in a reasonable amount of time. To select the interpolation mode, the downsampling quantization error was assessed using the spatial overlap (DICE score) between the original GT labels, and the GT labels undergoing 128-128-128 voxel downsampling, then upsampling back to the original size. Among the interpolation modes area, nearest and trilinear, the one giving the highest average DICE scores was trilinear, which were 94.8\% and 95.7\% for CT and US respectively. So, we used the trilinear interpolation mode for the downsampling.

As shown in \cite{reinhold2019evaluating, jacobsen2019analysis}, the intensity normalization of the input images, improves the performances of DL-based models processing images. So, we applied it to both 3DUS and CT images as a preprocessing step, by using the common approach consisting to subtract the mean channel's value from each input channel, then dividing the result by the channel standard deviation. That performance improvement by intensity normalization was also confirmed in our case since we have run some training/evaluation trials without it and obtained worse results. Standard data augmentation techniques were used during model training to circumvent overfitting, using randomized geometric transforms (flipping, rotation, translation, scaling, and affine distortion), and random intensity perturbation (intensity scaling, shifting, Gaussian noising, Gaussian smoothing, and contrast variation). 
% Full details of the augmentation parameters are provided in the supplementary material. 
The models nnUNet, CoTr and Glam were trained with the author implementations: Dice + Cross Entropy loss, SGD optimizer, a learning rate of $10^{-2}$, and weight decay of $10^{-5}$.
For 3D UNet and VNet, the framework Monai \cite{cardoso2022monai} was used with defaults training parameters, Dice + Cross Entropy loss, SGD optimizer, a learning rate of 1 with weight decay of $10^{-5}$, maximum number of epochs of 1000. Their training with a learning rate less than or equal to $10^{-1}$ and Dice + Cross Entropy loss, converged very slowly, hence the use of a learning rate equal to 1.

Post-inference, the predicted 128-128-128 voxel label maps were post-processed with connected component analysis to remove any spurious, and easily detectable, false positive regions.  Specifically, for CT, the label regions with the two largest connected components with respect to volume were retained, and for US, where only one kidney is visible at a time, the single largest connected component was retained. Finally, the label maps were upsampled with trilinear interpolation to the original image dimensions and then compared against the ground truth with several established performance metrics: DICE score, surface nearest neighbor distance (NN dist.), and $95\%$ Hausdorf Distance \cite{Dubuisson1994} (HD95 dist). NN dist. was computed using surface meshes that were automatically generated from the label maps using the Marching Cubes algorithm \cite{lewiner2003efficient}.
    
\subsubsection*{Quantitative Segmentation Results} 
\label{sec:quant_seg_results} 
\begin{table*}[!h]
    \begin{minipage}{1\linewidth}
    \centering
        \begin{tabular}{@{}c@{\hspace{7pt}}c@{\hspace{7pt}}|c@{\hspace{7pt}}c@{\hspace{7pt}}c@{\hspace{7pt}}|c@{\hspace{7pt}}c@{\hspace{7pt}}c@{\hspace{7pt}}c@{}}
        \toprule
              &          &  \multicolumn{3}{c}{CT} &  \multicolumn{3}{c}{US} \\
        \midrule
        Segmentation  & Training &  Dice score$\uparrow$ & NN dist.$\downarrow$ & HD95 dist.$\downarrow$ &  Dice score$\uparrow$ & NN dist.$\downarrow$ & HD95 dist.$\downarrow$ \\
         method  & target & in \% & in mm    & in  mm  & in \% & in mm    & in  mm   \\
        \midrule
        3DUNet & Double & 84.1 (0.8)* & 2.01 (0.62)* & 9.02 (3.27)* & 61.9 (4.4)* & 4.24 (0.56)* & 16.38 (2.93)*  \\ 
        3DUNet & Single & 83.7 (1.9)*  & 2.05 (0.64)* & 9.27 (3.14)* & 62.8 (3.1)* & 4.19 (0.55)* & 16.50 (2.61)*\\ 
        VNet    & Double & 87.3 (2.8)* & 1.70 (0.75)* & 7.55 (3.37)* & 72.8 (2.3)  * & 2.84 (0.63)* & 11.51 (3.00)* \\ 
        VNet    & Single & 88.3 (2.8)* & 1.57 (0.76)* & 6.82 (3.48)* & 73.7 (3.2)* & 2.69 (0.50) & 11.00 (2.50)*\\          
        nnUNet   & Double & 89 (1.5) & \textbf{1.26 (0.6)} & \textbf{5.49 (2.88)} & \textbf{79.4 (2.8)} & \textbf{2.29 (0.56)} & 9.96 (2.68) \\  
        nnUNet  & Single & 87.8 (5.6) & 1.74  (1.59) & 10.06 (12.19) & 79.0 (3.1) & 2.33 (0.37) & \textbf{9.59 (2.54)} \\  
        CoTr     & Double & 87.7 (1.6)* & 1.52 (0.67)* & 6.76 (3.25)* &  77.0 (2.2)* & 2.66 (0.56) & 11.47 (2.78) \\  
        CoTr     & Single & \textbf{89.1 (2.3)} & 1.38 (0.61)* & 5.91 (2.91)* & 76.5 (3.3) & 2.79 (0.85)*  & 11.95 (3.70)* \\  
        Glam     & Double & 84.9 (1.9)* & 1.85 (0.51)* & 7.54 (2.70) & 68.0 (11.6)* & 3.49 (1.97)* & 15.39 (7.51)* \\  
        Glam     & Single & 83.2 (5.9)* & 2.12 (1.30)* & 8.82 (4.61)* & 63.9 (12.2)* & 4.11 (1.91)* & 17.90 (7.27)* \\  
        \midrule
        Manual-Ann.1  & N/A & 95.7 (0.38) & 0.18 (0.01)  & 0.39 (0.03) &  94.45 (0.74) & 0.68 (0.12) & 1.91 (0.35)  \\ 
        Manual-Ann.2  & N/A & 95.4 (0.36) & 0.18 (0.01 )  & 0.39 (0.03) &  94.54 (0.39) & 0.67 (0.11) & 1.86 (0.33)  \\ 
        \bottomrule
        \end{tabular}
    \end{minipage}
    \caption{\footnotesize Quantitative evaluation of CT and US segmentation performance using the TRUSTED dataset. The ($\uparrow/\downarrow$) signs indicate whether higher or lower values are better. 5 automatic methods were compared (3DUNet, VNet, nnUNet, CoTr and Glam). Each method was trained and validated with 5-fold cross-validation (using the same splits), in two training configurations. The first configuration (Training target = 'Double') used two segmentation targets per training image (one from each annotator). The second configuration (Training target = 'Single') used one segmentation target per training image (the ground truth derived from the annotations using STAPLE). Performance was assessed on the held-out folds by comparing ground-truth with predictions (Dice, NN distance, and HD95 distance metrics). The table shows the mean performance averaged across all folds and the inter-fold standard deviation (bracketed). Bold indicates the automatic method with the best mean performance by metric. Stars indicate for each metric if there was a significant difference between a method's performance and the best performing method ($p<0.05$). The normal approximation of the Wilcoxon Signed Rank Test has been used for that. Performance of manual segmentation is presented at the bottom.}
    \label{tab:seg}
\end{table*}

Table \ref{tab:seg} compares the performances of the segmentation methods. The three metrics are presented for each model, and for each modality. Numbers represent inter-fold averages and standard deviations (bracketed).
In general nnUNet achieved the best performance: using double training targets, nnUNet obtained the best average values for 2 of 3 metrics, across both US and CT modalities. The worst performing method was 3D UNet for both CT and 3DUS modalities. US segmentation performance was generally lower compared to CT, with a difference of Dice score means, across all models, of approximately $15\%$, p-value = 0.5\% with the normal approximation of the Wilcoxon Signed Rank, indicating the significantly greater challenge associated to US kidney segmentation compared to CT. Moreover, there was a larger dispersion of segmentation performance for US compared to CT, indicated in all three metrics. The results show that state-of-the-art 3DUS kidney segmentation performance lags significantly behind that of CT, and more research is required to close the gap, through better architectures, learning setups, and/or hyper-parameter selection. The TRUSTED dataset will support such research efforts. Concerning single versus double training targets, their results were generally similar and there was no significant evidence to support the hypothesis that training on double targets, which captures annotation uncertainty information, improves segmentation performance. 

Table \ref{tab:seg} also shows the quantitative segmentation performance of each human annotator compared to GT, which were significantly better than the best automatic method (nnUNet). This indicates that on the TRUSTED dataset, there is still an important gap between human-level and state-of-the-art automatic segmentation. However, we acknowledge that because the GT labels were derived from the human annotations, the human-level performance may be artificially elevated.

\begin{figure*}[!h]
    \begin{minipage}{\linewidth}
    \centering
        \begin{tabular}{@{}c@{\hspace{1pt}}c@{\hspace{1pt}}c@{\hspace{1pt}}c@{\hspace{1pt}}c@{\hspace{1pt}}c@{\hspace{1pt}}c@{\hspace{1pt}}c@{\hspace{1pt}}@{}}
            \midrule
            & Manual Ann.1 & Manual Ann.2 & 3D UNet & VNet & nnUNet & CoTr & Glam \\
             \rotatebox{90}{\hspace{0.5pt} \scalebox{0.9}{CT KDY\_01}} & \includegraphics[width=0.1\textheight, height=0.1\linewidth]{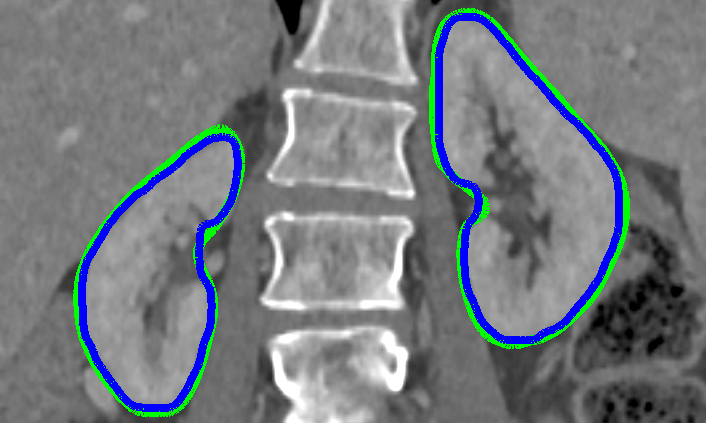} &
            \includegraphics[width=0.1\textheight, height=0.1\linewidth]{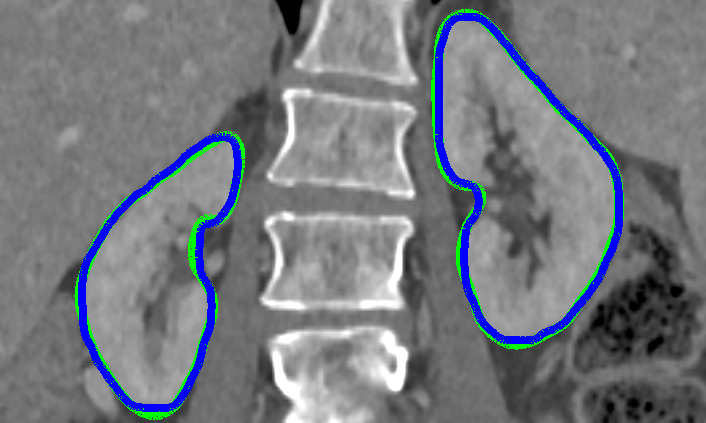} &
            \includegraphics[width=0.1\textheight, height=0.1\linewidth]{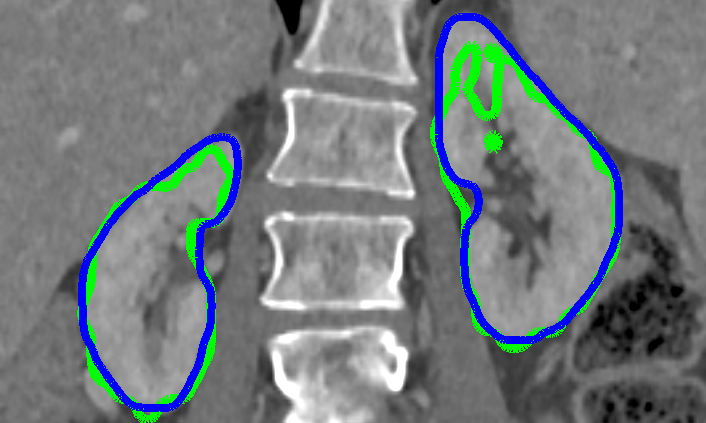} &
            \includegraphics[width=0.1\textheight, height=0.1\linewidth]{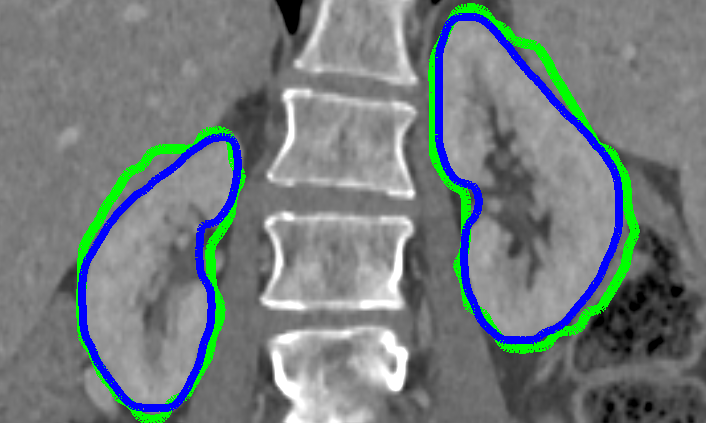} &
            \includegraphics[width=0.1\textheight, height=0.1\linewidth]{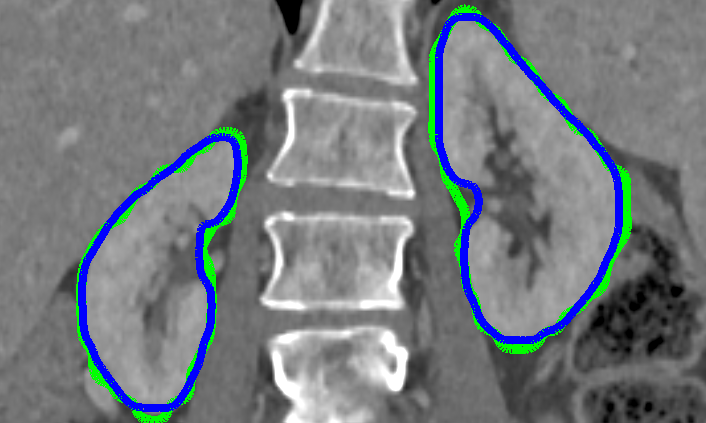} &    
            \includegraphics[width=0.1\textheight, height=0.1\linewidth]{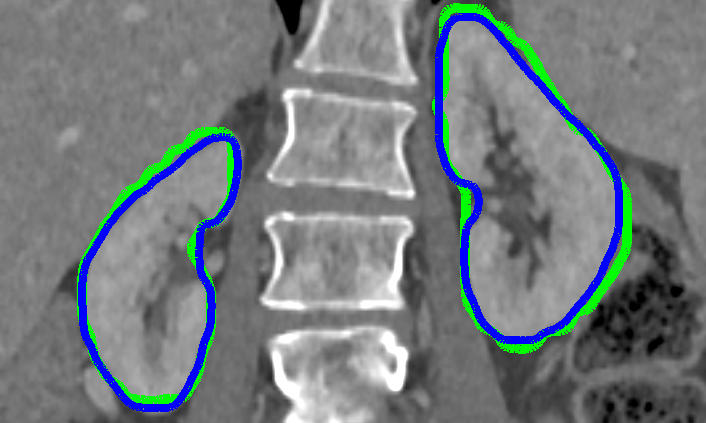} &
            \includegraphics[width=0.1\textheight, height=0.1\linewidth]{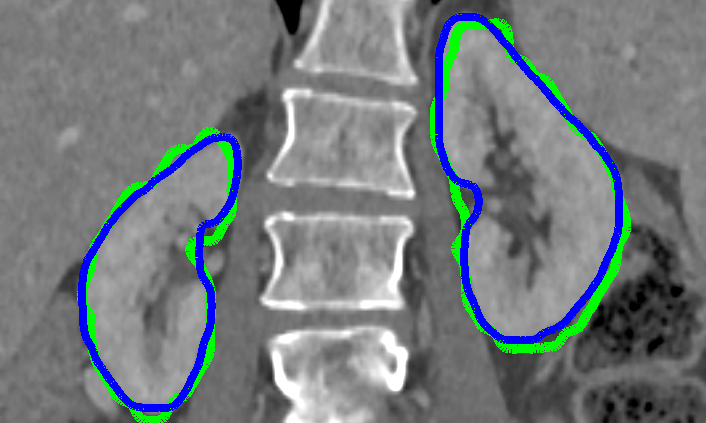}\\
            
            \rotatebox{90}{\hspace{0.5pt} \scalebox{0.9}{CT KDY\_17}} & \includegraphics[width=0.1\textheight, height=0.1\linewidth]{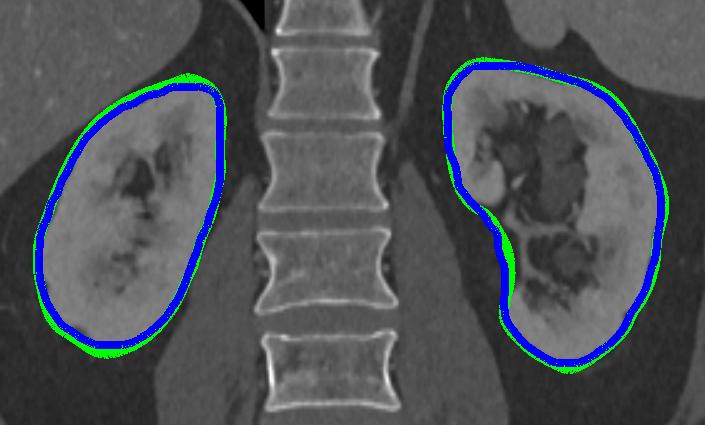} &
            \includegraphics[width=0.1\textheight, height=0.1\linewidth]{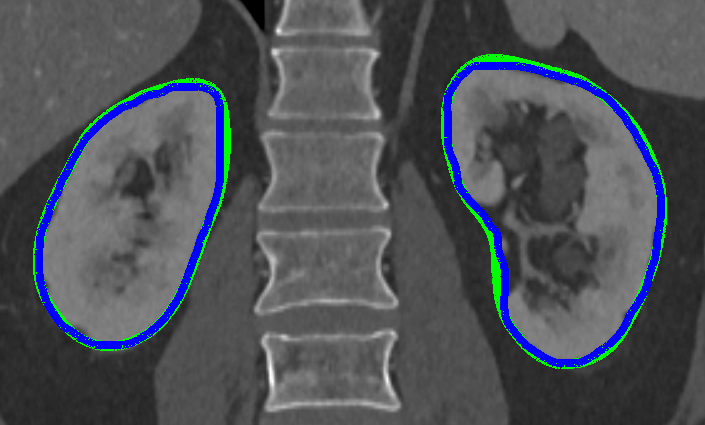} &
            \includegraphics[width=0.1\textheight, height=0.1\linewidth]{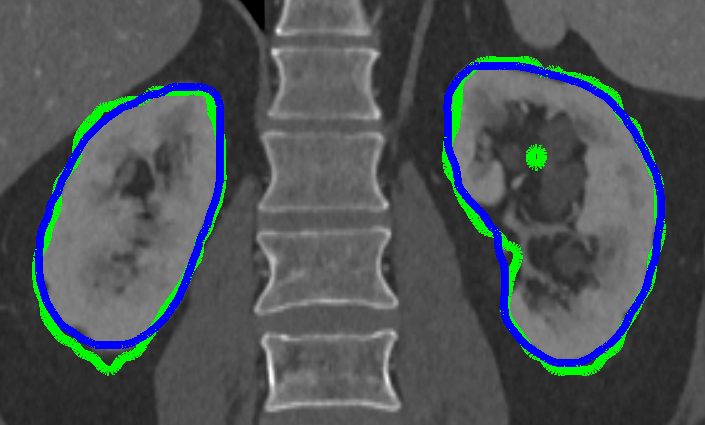} &
            \includegraphics[width=0.1\textheight, height=0.1\linewidth]{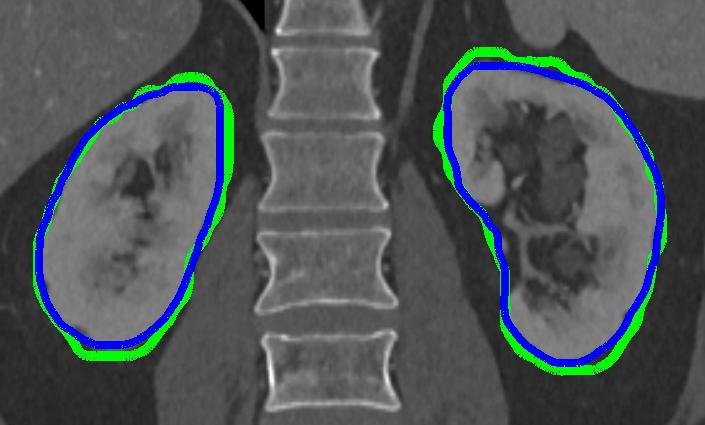} &
            \includegraphics[width=0.1\textheight, height=0.1\linewidth]{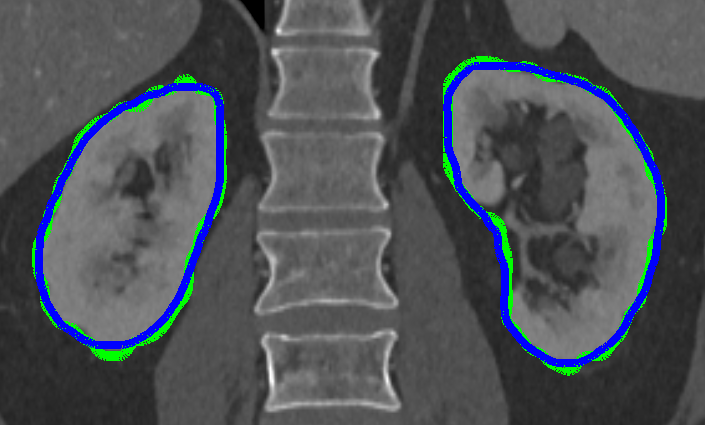} &    
            \includegraphics[width=0.1\textheight, height=0.1\linewidth]{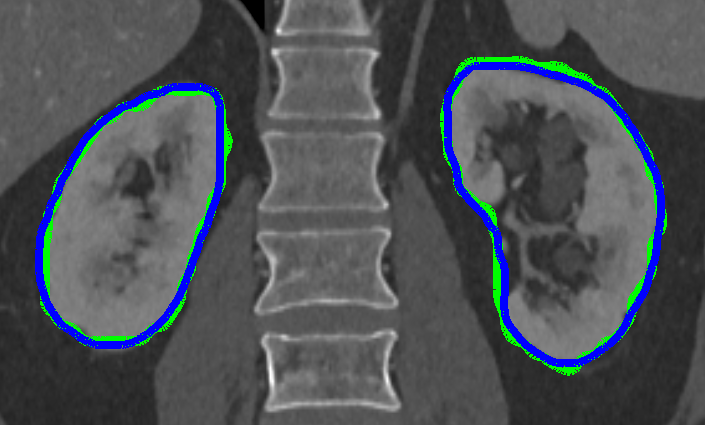} &
            \includegraphics[width=0.1\textheight, height=0.1\linewidth]{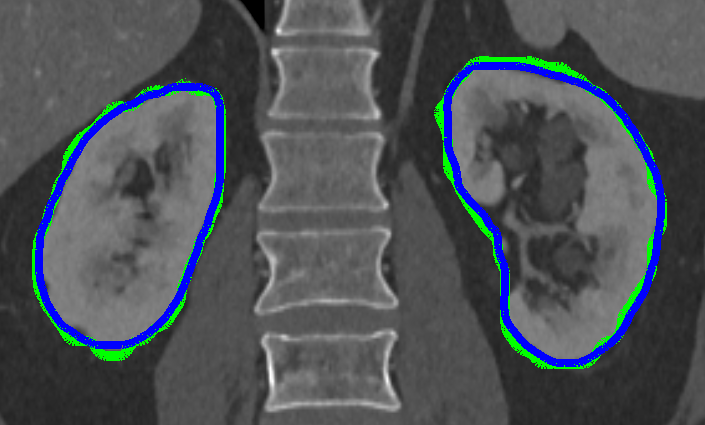}\\

            \midrule

            \rotatebox{90}{\scalebox{0.9}{US KDY\_01R}} & \includegraphics[width=0.1\textheight, height=0.1\linewidth]{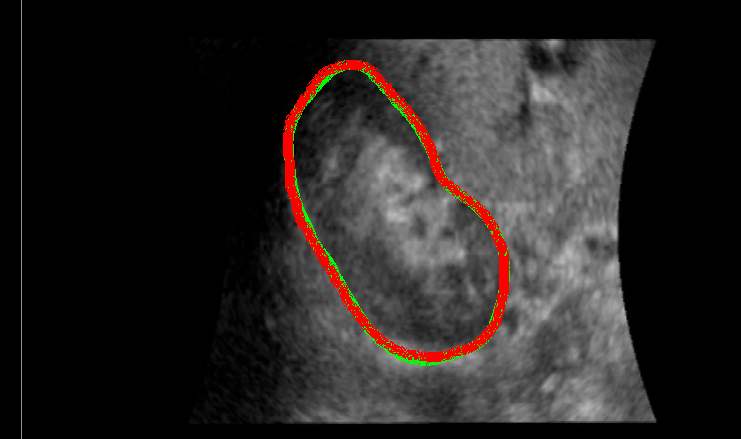} &
            \includegraphics[width=0.1\textheight, height=0.1\linewidth]{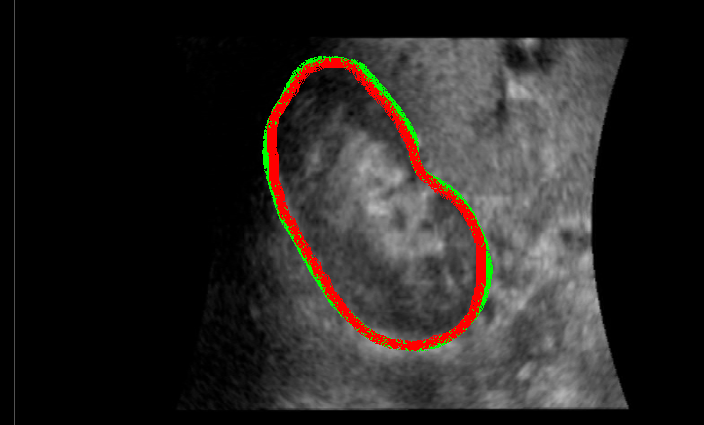} &
            \includegraphics[width=0.1\textheight, height=0.1\linewidth]{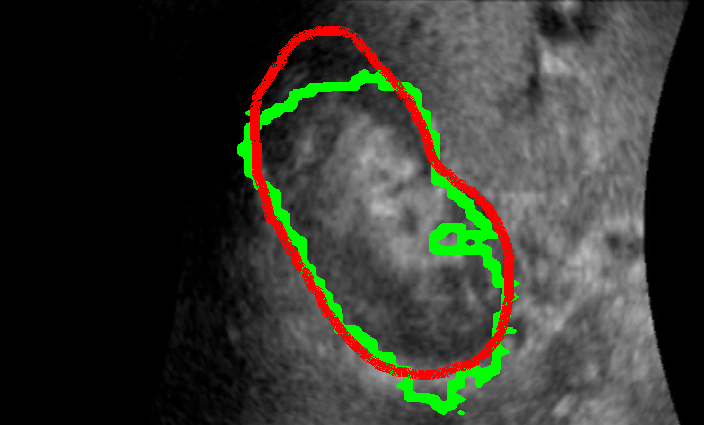} &
            \includegraphics[width=0.1\textheight, height=0.1\linewidth]{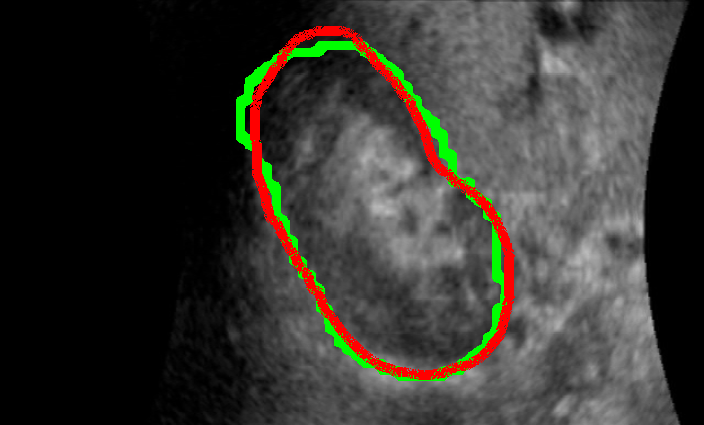} &
            \includegraphics[width=0.1\textheight, height=0.1\linewidth]{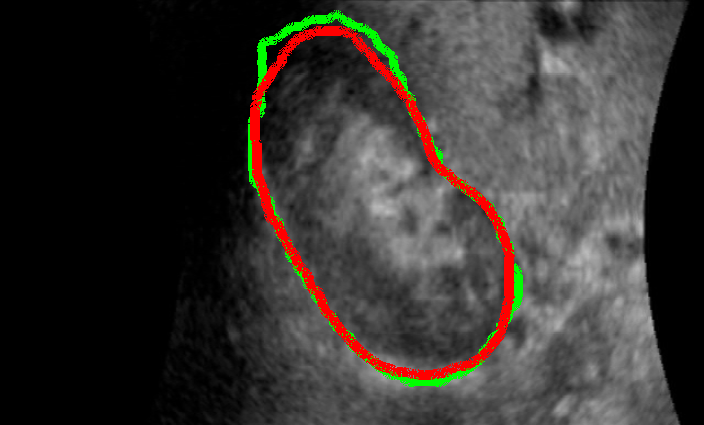} &    
            \includegraphics[width=0.1\textheight, height=0.1\linewidth]{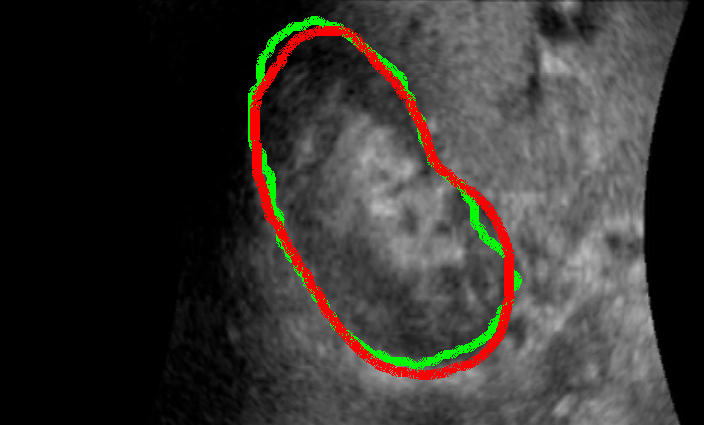} &
            \includegraphics[width=0.1\textheight, height=0.1\linewidth]{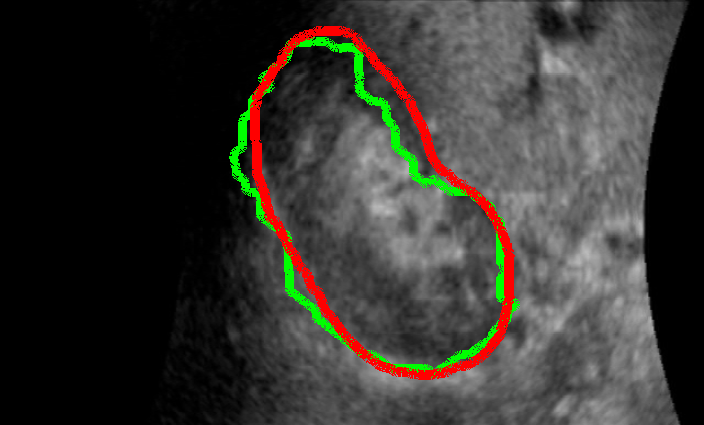}\\

            \rotatebox{90}{\scalebox{0.9}{US KDY\_17L}} & \includegraphics[width=0.1\textheight, height=0.1\linewidth]{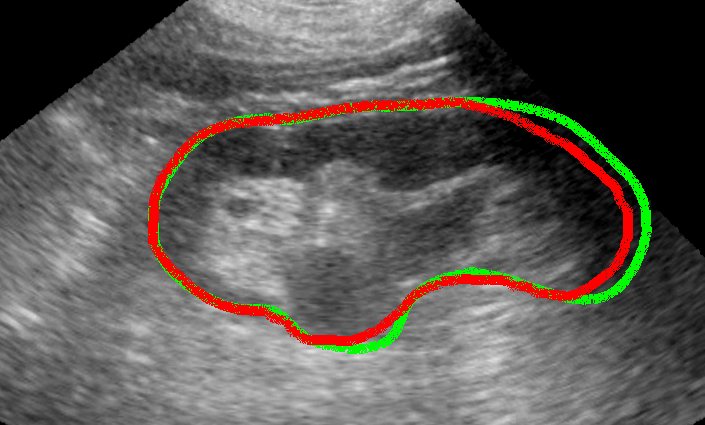} &
            \includegraphics[width=0.1\textheight, height=0.1\linewidth]{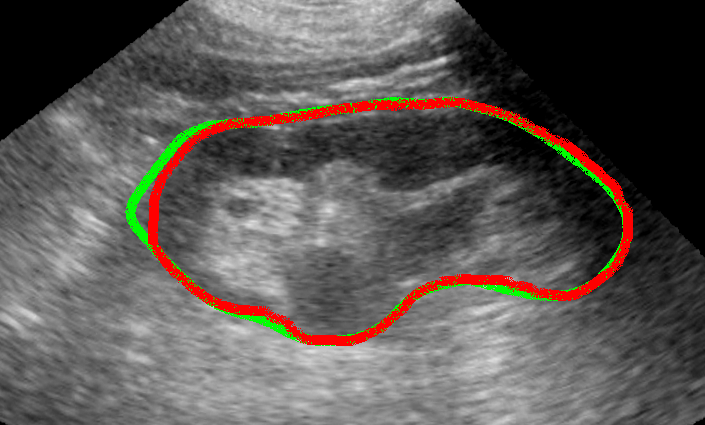} &
            \includegraphics[width=0.1\textheight, height=0.1\linewidth]{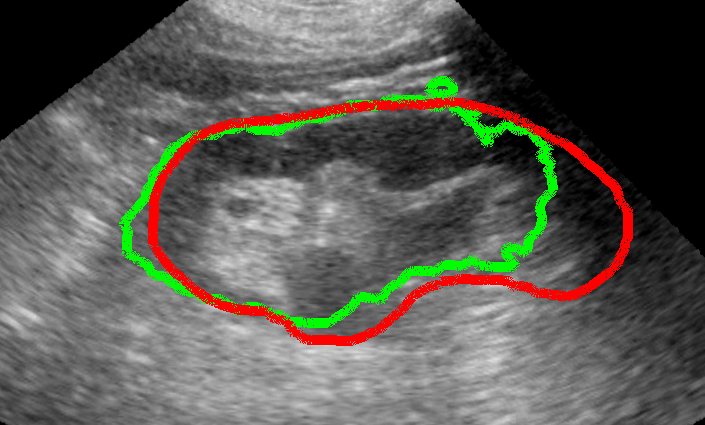} &
            \includegraphics[width=0.1\textheight, height=0.1\linewidth]{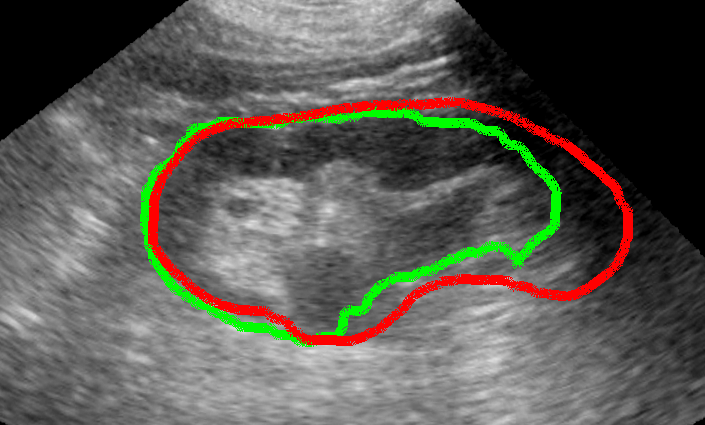} &
            \includegraphics[width=0.1\textheight, height=0.1\linewidth]{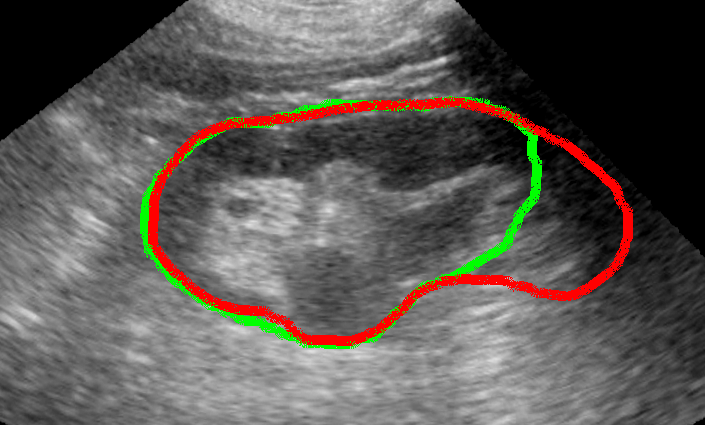} &    
            \includegraphics[width=0.1\textheight, height=0.1\linewidth]{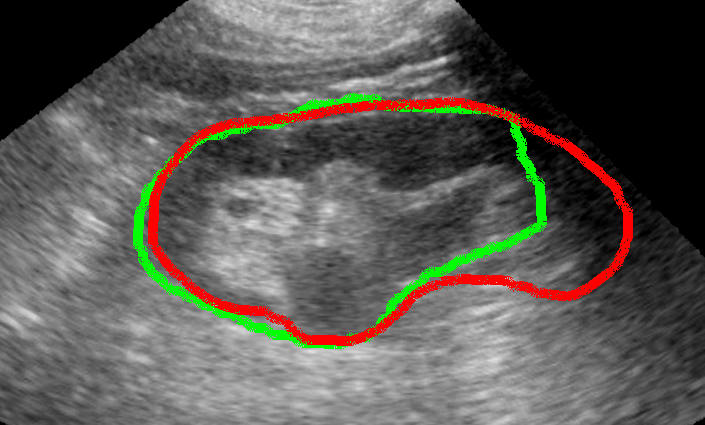} &
            \includegraphics[width=0.1\textheight, height=0.1\linewidth]{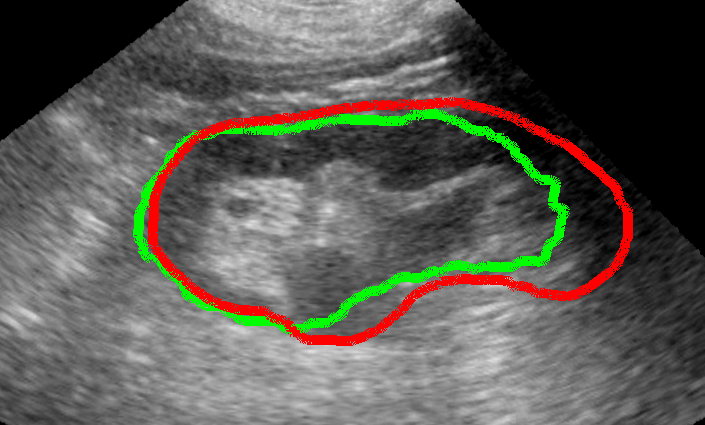}\\
        \end{tabular}

    \begin{tabular}{ccc}
         {\color{blue}{\line(0,1){2}}}{\quad \footnotesize Kidney ground-truth segmentation in CT, } &
         {\color{red}{\line(0,1){2}}}{\quad \footnotesize Kidney ground-truth segmentation in US, }  &
         {\color{green}{\line(0,1){2}}}{\quad \footnotesize Kidney segmentation with different methods. } 
    \end{tabular}

    \end{minipage}
    \caption{\footnotesize Qualitative results showing example CT and US segmentations using the TRUSTED dataset (patient 01 and 17). The top two rows shows coronal CT slices, with ground-truth segmentations overlaid in blue, and estimated segmentations in green. The bottom two rows shows longitudinal US slices, with ground truth segmentations overlaid in red and estimated segmentations in green. The two first columns (Manual Ann.1 and Manual Ann.2) show segmentations from each annotator, and the remaining columns show the best training version on average between single or double target(s) of each automatic segmentation from 5 DNN-based methods.}
    \label{fig:USseg}
\end{figure*}

\subsubsection*{Qualitative Segmentation Results} 
Figure \ref{fig:USseg} shows results of automatic CT and 3DUS kidney segmentation with two patient examples (01, and 17). Each figure column corresponds to a segmentation method (the first two columss are manual segmentation from either annotator, and the remaining are the automatic methods). Ground truth CT and US segmentations are overlaid as blue and red contours respectively, and segmentations from each method are shown in green. For the automatic methods, results of the best training configuration (single versus double targets) are shown (with the highest average inter-fold DICE score). Considering the manual segmentations, one can observe very close agreement in CT and in US kidney 01R, but some stronger disagreement in US kidney 17L. This disagreement (US KDY\_17L) is due to the poorer image contrast at the right kidney border caused by a rib shadow. Nevertheless, the manual segmentations are relatively close with a DICE overlap of 84\%.

The close performance of the automatic methods for CT segmentation, as reported in Table \ref{tab:seg} are reflected in Figure \ref{fig:USseg}. In contrast, performance for US segmentation is more varied. Generally, US segmentation performance is worse at the shadow region of case Kidney 17L, where there is much poorer image contrast, required to precisely delineate the organ. For this case, all automatic methods under-segment the right region to some extend, the best methods are nnUNet and Glam with the lowest under-segmentation. Notice that neither nnUNet nor Glam systematically under-segment, as shown in Case Kidney 01R.

\subsection*{Inter-Modal Image Registration}
\subsubsection*{Methods and Evaluation Methodology}
In this section, three competitive IMIR methods are compared using the TRUSTED dataset. The most accurate registration methods use a refinement process, where a rough initial registration is provided, which is subsequently improved by the method. The first method is Iterative Closest Points (ICP) \cite{icp1987}, which despite its age, remains a highly competitive surface-based method regarding both registration accuracy and speed \cite{du2020robust}. The second method is Bayesian Coherent Point Drift (BCPD) \cite{hirose2021}, which improves on the popular Coherent Point Drift (CPD) method in optimization speed and a wider registration convergence basin. Both ICP and BCPD require surface segmentations, which were generated by nnUNet trained with double targets (based on its strong performance in both modalities as described in the Automatic Segmentation section. Surfaces were generated by Lewiner's version of the Marching Cubes algorithm \cite{lewiner2003efficient}. %Due to space restrictions, we did not include an analysis of registration performance using different segmentation methods. 

The third registration method is from ImFusion Suite 2.45.3 (ImFusion GmbH, Munich, Germany), which is a state-of-the-art intensity-based method using iterative optimization, specifically designed for intermodal 3DUS registration \cite{horstmann2022orientation, markova2022global}. Two variants have been compared using different image similarity metrics: Local Normalized Cross Correlation (IF-LNCC), and Local Cross-Correlation (IF-LC2).

All registration methods require a spatial transform model. Based on various previous works that show that physiological movement of the kidney does not induce significant nonlinear deformation  \cite{Ong2009, figueroa2014biomechanical, wein2008automatic}, we compared two transform models: rigid motion, with 6 degrees-of-freedom (DOFs), and affine motion, with 12 Dofs. Default hyper-parameters were used for ICP, BCPD, IF-LNCC and IF-LC2.
% (... which are provided in the supplementary material)

\subsubsection*{Registration Initialisation}
For a fair comparison, all registration methods were initialised with the same landmark-based approach using 3 landmarks: \#3: center-lateral, \#4: superior-pole and \#5: posterior-pole, see Figure \ref{fig:manualann}, and the best-fitting 3D similarity transform in the least-squares sense was used to provide the initial registration. Sensitivity to the initial registration accuracy was assessed with a semi-synthetic experiment where the landmark positions in 3DUS image coordinates were randomly perturbed by varying amounts of white Gaussian noise, to simulate varying initial registration imprecision. This experiment is presented in the Quantitative Registration Results section.

\subsubsection*{Evaluation Metrics}
Four complementary metrics were used to evaluate the registration method accuracy:
\begin{itemize}
        \item \textbf{Target Registration Error} (TRE), which evaluates the distance between the GT position of a target landmark (one not used for registration), and its predicted 3D position according to the estimated registration. We used the renal pelvis centre (landmark \#1) as the target, which was the only internal registration landmark that could be reliably located in both CT and US coordinates by the annotators. Recall that landmark \#1 annotations had a mean distance to GT of 1.82 mm in the CT images, and 2.23 mm in the 3DUS images. 
        \item \textbf{Dice score}, which evaluates the spatial overlap between the GT CT segmentation and the GT 3DUS segmentation transformed according to the estimated registration.
        \item  \textbf{Nearest Neighbour surface distance} (NN dist.), which evaluates the average nearest neighbour distance between the GT CT segmentation surface and the GT 3DUS segmentation surface transformed according to the estimated registration.  
        \item \textbf{Hausdorff 95 distance} (HD95), which evaluates the surface distance between the GT CT segmentation and the registered GT US segmentation according to the $95\%$ Hausdorff distance.
\end{itemize}

\subsubsection*{Quantitative Registration Results} 
\label{sec:quant_reg_results}
Table \ref{tab:reg} shows the quantitative performances of the registration refinement methods. We refer the reader to the table caption for a thorough description of the table components. In the table's top horizontal section (showing the performance of methods that can be used in practice), BCPD using affine transforms obtains the best mean performance metrics in general, with a mean TRE of 4.53 mm and mean Dice score of 84.16\%. This is followed by BCPD with a rigid transform, with a mean TRE of 4.90 mm and mean Dice score of 82.08\%. The relatively small improvement can be partially explained by the fact that the initial registration uses a similarity transform, which already accounts for some non-rigid effects (isotropic scaling). The surface-based methods (both ICP and BCPD) generally outperform the intensity-based ones (IF-LNCC and IF-LC2), with lower mean performance metrics across all metrics. This shows two aspects. Firstly, the intensity-based methods were not able to exploit potential extra registration information contained in the voxel intensity values, Secondly, the worse performance indicated the intensity-based methods were struggling to converge on the correct solution, compared to the surface-based methods.

Considering the bottom section of Table \ref{tab:reg} (showing surface-based registration performance using GT surfaces), BCPD using affine transforms provided the best results across all metrics, and additionally, the performance exceeded that of BCPD using automatic segmentation (from nnUNet): The mean TRE was 4.13 mm and mean Dice score was $89\%$. The results are in line with \cite{leroy2007intensity}, which evaluated ICP performance using manual segmentation on a phantom dataset (obtaining a mean TRE of 4.7 mm). The improves results with manual segmentation on this human dataset motivates the need for improved automatic segmentation methods, to reach (or surpass) registration performance with manual segmentation. Consequently, this dataset can be used to assess and improve automatic segmentation methods considering a concrete downstream application (registration). 

\begin{table*}[!h]
    \begin{minipage}{\linewidth}
    \centering
    \begin{adjustbox}{max width=\textwidth}
        \begin{tabular}
        {@{}c@{\hspace{4pt}}c@{\hspace{4pt}}|c@{\hspace{4pt}}c@{\hspace{4pt}}c@{\hspace{4pt}}c@{\hspace{4pt}}|c@{\hspace{4pt}}c@{\hspace{4pt}}c@{\hspace{4pt}}c@{}}
        \toprule
               &          &  \multicolumn{4}{c}{\textbf{Rigid transform assumption}} & \multicolumn{4}{c}{\textbf{Affine transform assumption}}\\
        \midrule
         Registration  &  Segmentation & TRE$\downarrow$ & Dice score $\uparrow$ & NN dist. $\downarrow$ &  HD95 dist.$\downarrow$ & TRE$\downarrow$ & Dice score $\uparrow$ & NN dist.$\downarrow$ &  HD95 dist.$\downarrow$\\
         refine method & method &  in mm  &  in \%  &  in mm &  in mm&  in mm  &  in \%  &  in mm &  in mm\\
        \midrule
         ICP  (S) & nnUNet &  5.37 (0.7)* & 80.30 (1.34)*  & 2.47 (0.23)* & 8.61 (1.06)* &  5.42 (0.88)* & 80.60 (1.30)*  & 2.51 (0.25)* & 8.56 (1.24)* \\
         BCPD (S) & nnUNet  & 4.90 (0.64)* & 82.08 (2.28)*  & 2.35 (0.22) & 7.92 (1.20)* & \textbf{4.53 (0.79)} & \textbf{84.16 (1.85)}  & \textbf{2.19 (0.41)} & \textbf{7.32 (1.28)} \\
         IF-LNCC (I) & N/A &  7.86 (2.19)* & 72.32 (4.84)*  & 2.97 (0.34)* & 12.70 (3.11)* &  6.13 (0.92)* & 74.89 (1.58)* & 3.08 (0.46)* & 11.24 (1.09)* \\
         IF-LC2 (I) & N/A &  6.37 (1.62) & 74.87 (4.00)*  & 2.65 (0.21)* & 10.83 (1.63)* &  6.16 (1.53) & 75.77 (3.23)*  & 2.80 (0.33)* & 10.79 (1.51)* \\
         \midrule
         ICP (S) & STAPLE-GT &  4.64 (0.69) & 85.16 (1.31)  & 2.20 (0.21) & 6.50 (0.83) &  4.78 (0.64) & 87.17 (1.62)  & 1.86 (0.26) & 5.87 (0.87)\\
         BCPD (S) & STAPLE-GT &  4.69 (0.58) & 85.28 (1.27)  & 2.21 (0.20) & 6.62 (0.83) &  4.10 (0.43) & 89.03 (1.13)  & 1.50 (0.21) & 5.18 (0.72)\\
        \bottomrule
        \end{tabular}
    \end{adjustbox}
    \end{minipage}
    \caption{\footnotesize Quantitative results of registration refinement methods using the TRUSTED dataset. The ($\uparrow/\downarrow$) signs indicate whether higher or lower values are better. The table is divided into two main parts: Left - results of methods configured to use rigid spatial transforms (6DoF), Right - results of methods configured using affine spatial transforms (12 DoF). All registration refinement methods were initialised with the same spatial transforms (similarity transform computed from manually-annotated landmarks \# 3, \# 4 and \# 5). For the surface-based registration methods indicated by '(S)', the method to generate surface segmentations for CT and US modalities is given in column 2. Columns 3-10 give the mean and standard deviation (bracketed) for each of the 4 performance metrics (5-fold averaging). For each metric, bold items show the best performing method (excluding surface-based methods using GT segmentations). Stars indicate for each metric if there was a significant difference between a method's performance and the best performing method ($p<0.05$). The normal approximation of the Wilcoxon Signed Rank Test has been used for that.}
    \label{tab:reg}   
\end{table*}

In clinical context, a registration refinement method that require a precise initial registration is not desirable. Consequently, we conducted an initialization sensitivity analysis using the TRUSTED dataset to quantify the effect of initial registration accuracy on method performance. For each CT/3DUS image pair, random white noise was applied to the three registration landmarks with a standard deviation $\sigma$ ranging from 0 mm (no noise) to 10 mm (strong noise) in 2 mm increments. For each $\sigma$, registration was initialized using the best-fitting similarity transform using the noisy landmarks, and then each registration refinement method was run. To generate different noise samples, the process was repeated 5 times for each noise level. In total 14,160 registrations have been performed: 4 registration methods, 2 spatial transform models, 6 noise levels, 5 noise samples, 59 image pairs).

\begin{figure*}[!h]
    \begin{minipage}{1\linewidth}
    \centering
    \begin{tabular}{c}
        \includegraphics[width=0.97\linewidth]{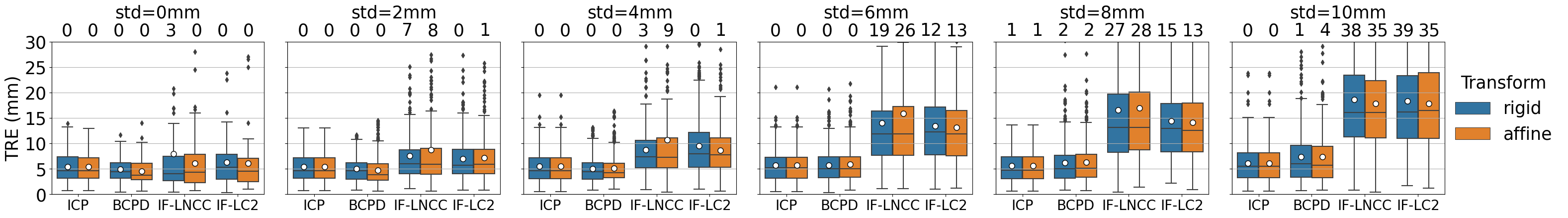} \\
        \footnotesize a - Target registration Error (TRE)\\
        \midrule
        \includegraphics[width=0.97\linewidth]{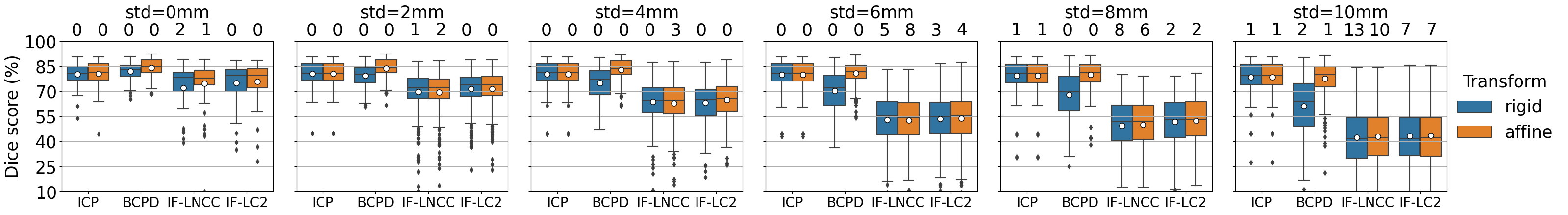} \\
        {\footnotesize b - Dice score}\\
        \midrule
        \includegraphics[width=0.97\linewidth]{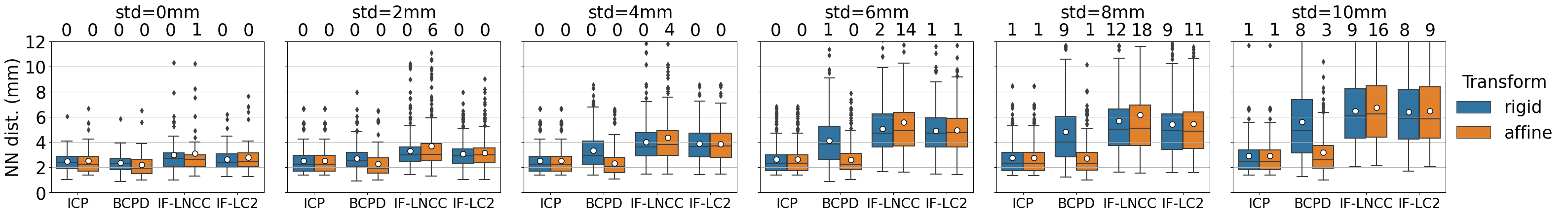} \\
        {\footnotesize c - Nearest neighbor (NN) distance}\\
        \midrule
        \includegraphics[width=0.97\linewidth]{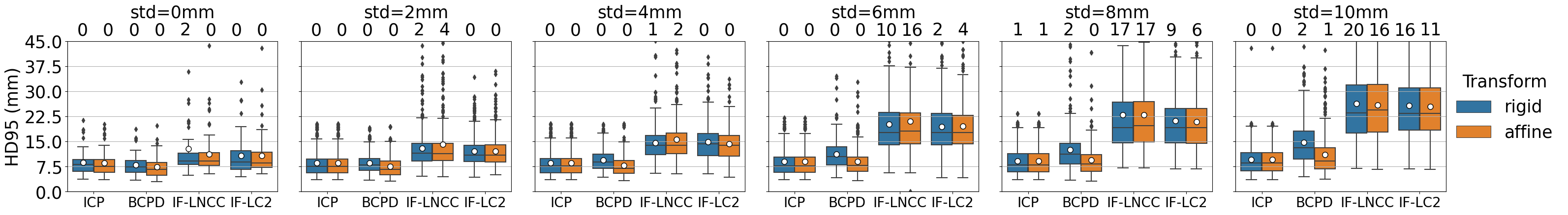} \\
        {\footnotesize d - H95 distance}\\

        \bottomrule
    \end{tabular}
    \end{minipage}
    \caption{\footnotesize Initialization sensitivity trends of registration refinement methods using the TRUSTED dataset. The figure is divided into 4 rows (one row per performance metric). Each row has 6 sub-figures, showing the performance of every compared method using a different amount of artificial noise added to the initial registrations (ranging from 0 mm (no noise) to 10 mm std (large noise). Boxplots are used to summarize the performance distribution of each method in two configurations (using rigid and affine spatial transforms). For each boxplot, the lower, central and upper bars give the 1st, 2nd (median) and 3rd quartiles. White circles give the mean, and whiskers give the extreme values not considered as outliers. Dots give outlier values, and the numbers above each boxplot give the number of outliers that are beyond the $y$-axis range.}
    \label{fig:regbox}
\end{figure*}

The results are presented in Figure \ref{fig:regbox} (please refer to the caption for a details description of the figure layout). One can see a general degradation of registration performance with increased $\sigma$ (initial registration noise). When $\sigma=0$, the results are identical to those presented in Table \ref{tab:reg} but presented in a box-plot format. The median Dice scores are all over 75\%, and the median TRE, NN dist. and HD95 dist. are all under 5 mm, 3 mm and 10 mm respectively. The best performing method is BCPD, agreeing with Table \ref{tab:reg}. However, all methods have strong tails, indicating that an important minority of image pairs have high errors. The 3rd ($75th\%$) quartle TRE values (\textit{i.e.} the box top edges) are above 5 mm for all methods. This indicates that even if the registration methods are provided with a relatively good initialization (using manual landmarks), the TRUSTED dataset brings a methodological challenge to reduce the proportion of cases with high registration errors. 

As registration initialization noise increases, the performance of the intensity-based methods (IF-LNCC and IF-LC2) degrades more strongly compared to the surface-based methods. For BCPD, the affine motion model clearly produces better results for Dice score, NN Dist. and HD95 dist, compared to rigid motion, however, the the difference is small when considering TRE. We believe the reason for this is that the location of the landmark used for TRE (renal pelvis) is often close to the organ's centroid, and consequently, registration of this point with rigid and affine models generally produces a similar moved target. 

Nevertheless, as noise increases all methods have an increased proportion of poorly registered image pairs, which clearly shows the strong important that initialisation accuracy has on final registration error. The same trends are apparent for all 4 metrics. The results indicate that there is a need to improve the all methods, and particularly the best performing methods to reduce sensitivity to high initialisation noise.     

\subsubsection*{Qualitative Registration Results}

\begin{figure*}[!h]
% \centering
    \begin{minipage}{\linewidth}
    \centering
        \begin{tabular}{@{}c@{\hspace{1pt}}c@{\hspace{1pt}}c@{\hspace{1pt}}c@{\hspace{1pt}}c@{\hspace{1pt}}c@{\hspace{1pt}}@{}}

            & ICP & BCPD & IF-LNCC & IF-LC2 \\
            \midrule
            \rotatebox{90}{\hspace{6pt}KDY\_01R, Rigid} & \includegraphics[height=0.15\textwidth, width=0.15\textheight]{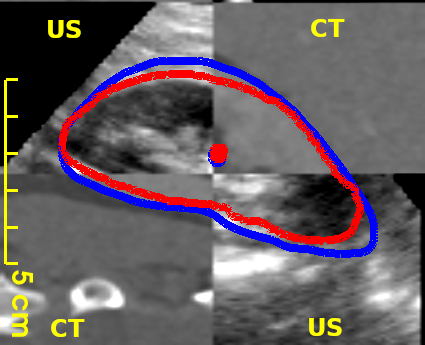} &
            \includegraphics[height=0.15\textwidth, width=0.15\textheight]{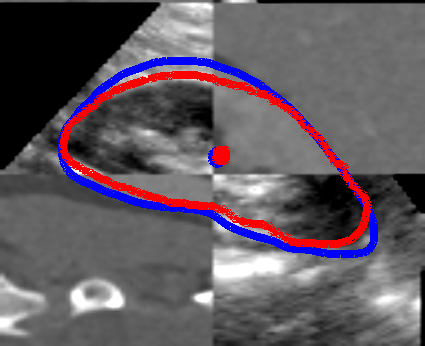} &
            \includegraphics[height=0.15\textwidth, width=0.15\textheight]{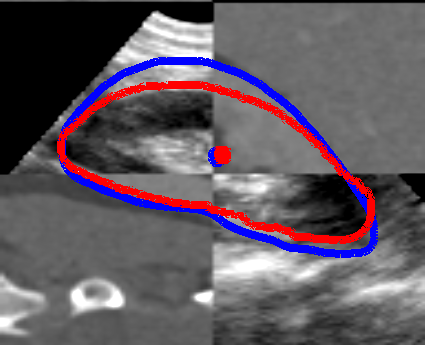} &
            \includegraphics[height=0.15\textwidth, width=0.15\textheight]{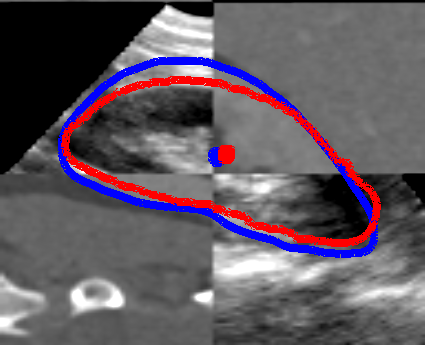}\\
            
            \rotatebox{90}{\hspace{4pt}KDY\_01R, Affine} & \includegraphics[height=0.15\textwidth, width=0.15\textheight]{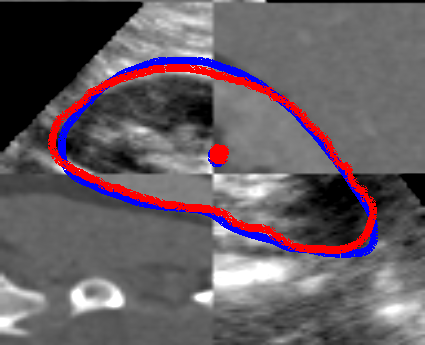} &
            \includegraphics[height=0.15\textwidth, width=0.15\textheight]{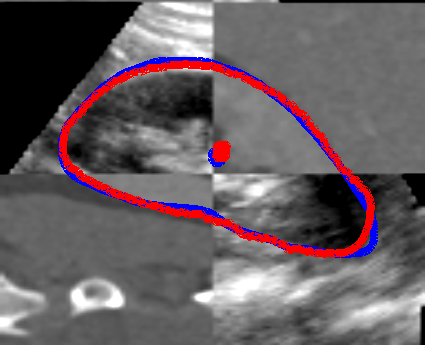} &
            \includegraphics[height=0.15\textwidth, width=0.15\textheight]{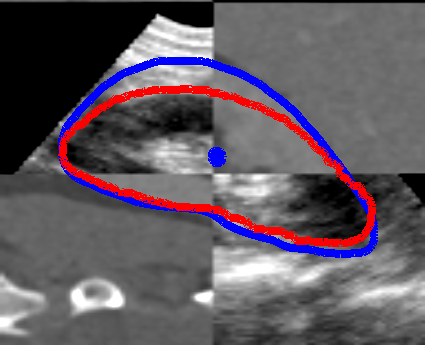} &
            \includegraphics[height=0.15\textwidth, width=0.15\textheight]{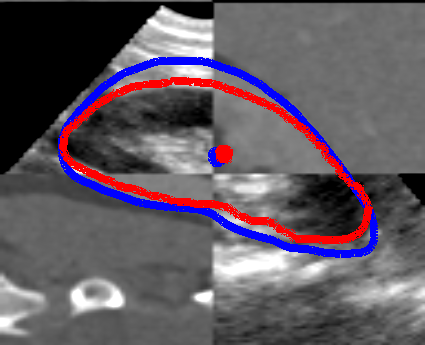}\\

            \midrule
            \midrule

            \rotatebox{90}{\hspace{6pt}KDY\_17L, Rigid} & \includegraphics[height=0.15\textwidth, width=0.15\textheight]{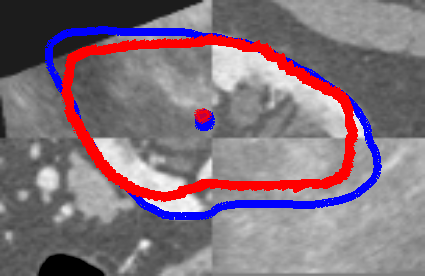} &
            \includegraphics[height=0.15\textwidth, width=0.15\textheight]{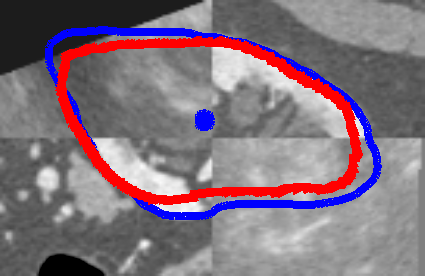} &
            \includegraphics[height=0.15\textwidth, width=0.15\textheight]{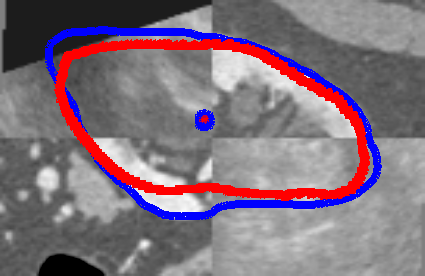} &
            \includegraphics[height=0.15\textwidth, width=0.15\textheight]{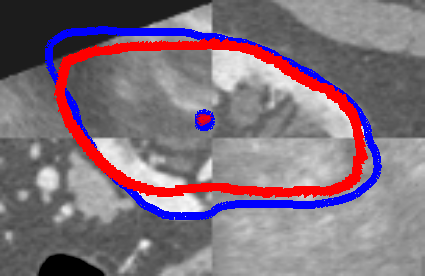}\\

            \rotatebox{90}{\hspace{4pt}KDY\_17L, Affine} & \includegraphics[height=0.15\textwidth, width=0.15\textheight]{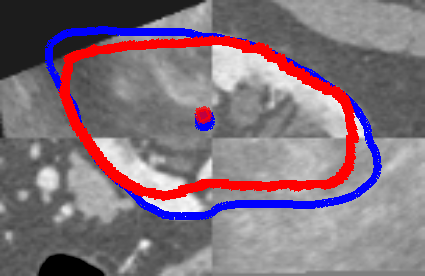} &
            \includegraphics[height=0.15\textwidth, width=0.15\textheight]{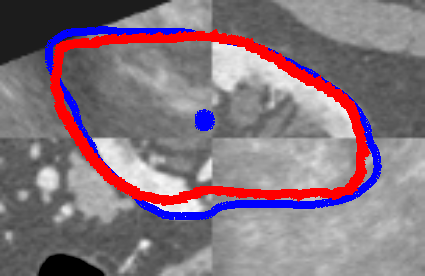} &
            \includegraphics[height=0.15\textwidth, width=0.15\textheight]{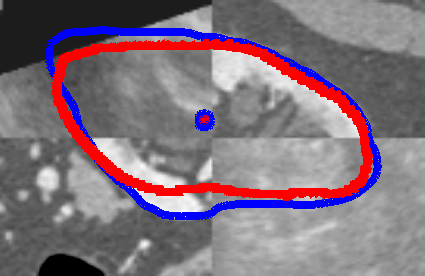} &
            \includegraphics[height=0.15\textwidth, width=0.15\textheight]{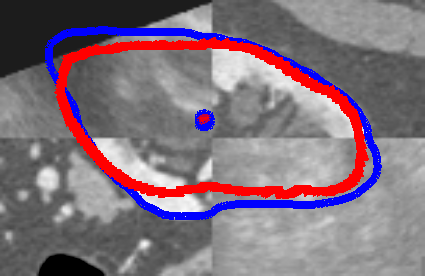}
        \end{tabular}
        
    \begin{tabular}{cc}
         {\color{blue}{\line(0, 1){2}}}{\quad \footnotesize Kidney ground-truth annotation in CT } &
         {\color{red}{\line(0, 1){2}}}{\quad \footnotesize Kidney ground-truth annotation in US } 
    \end{tabular}

    \end{minipage}
    \caption{\footnotesize Qualitative results showing registration performance using the TRUSTED dataset. Each figure column corresponds to a method. The first and second figure rows show results with one example (KDY\_01R), using a rigid and affine registration refinement transform. The third and fourth figure rows show results with a second example (KDY\_17L), using a rigid and affine registration refinement transform. Each image in the figure shows 4 annotations: Blue contour - the GT kidney segmentation in CT. Blue point - the GT position of Landmark 1 in CT. Red contour - the GT kidney segmentation in US, moved to CT coordinates according to the estimated registration. Red point - the GT position of Landmark 1 in US, moved to CT coordinates according to the estimated registration. A checkerboard visualization is used to show the fixed image (CT) overlaid with the registered image (US).}
    \label{fig:regcheck}
\end{figure*}

Figure \ref{fig:regcheck} shows visual registration results with $\sigma=0$, using the the same two cases as Figure \ref{fig:USseg} (KDY$\_$01R: Patient 01 - right kidney, and KDY$\_$17L: Patient 17, left kidney). For ICP and BCPD, automatic US and CT segmentations from nnUNet trained with double targets were provided as inputs. Please see the figure legend for a detailed explanation of the visualisations. For KDY$\_$01R, all registration methods perform relatively well, however ICP and BCPD show a better alignment of the GT surfaces (which we recall, were not used by these methods), compared to IF-LNCC and IF-LC2. For KDY$\_$17L, registration accuracy was generally worse than KDY$\_$01R, mainly due to a strong shadow that significantly reduced the contrast of the kidney's borders. While knowing that, visualizing a single plan is not enough to appreciate the volumes overlapping, in the case KDY$\_$17L, IF-LC2 shows a marginally better result than IF-LNCC and BCPD, on this plane. One can also see that it is difficult to conclude about the quality registration just by using the TRE metric. Therefore, the GT segmentation alignment metrics (DICE, NN dist., and HD95 dist.) are important metrics to consider, in addition to TRE, when evaluating methods on this dataset.

\section*{Usage Notes}
Before downloading the dataset, users must first agree and email a signed copy of the user licence agreement as mentioned in the Data Records section. Users should make sure to adhere to the licence agreement when using this dataset. If in doubt, they should contact IRCAD France any queries. Nifti images were used to store image and segmentation annotations, since it can easily be used in well-known frameworks such as Monai \cite{cardoso2022monai}. The splits used for 5-fold cross validation are included to allow equivalent comparisons with the results presented in this paper.

\section*{Code Availability}
Due to intellectual property restrictions, the code used to train models and generate performance data is not publicly available. 
%Here, for reproducibility, we list all parameters used to train the DL models, and details of the code libraries and library versions used to train models and generate results.
%\begin{itemize}
%    \item setting table shared with Loic
%    \item the dependencies (packages) used
%    \item versions of software used
%\end{itemize}

\bibliography{trusted_datapaper}

\begin{thebibliography}{10}
\urlstyle{rm}
\expandafter\ifx\csname url\endcsname\relax
  \def\url#1{\texttt{#1}}\fi
\expandafter\ifx\csname urlprefix\endcsname\relax\def\urlprefix{URL }\fi
\expandafter\ifx\csname doiprefix\endcsname\relax\def\doiprefix{DOI: }\fi
\providecommand{\bibinfo}[2]{#2}
\providecommand{\eprint}[2][]{\url{#2}}

\bibitem{european2019abdominal}
\bibinfo{author}{ESR}.
\newblock \bibinfo{journal}{\bibinfo{title}{Abdominal applications of
  ultrasound fusion imaging technique: liver, kidney, and pancreas}}.
\newblock {\emph{\JournalTitle{Insights into Imaging}}}
  \textbf{\bibinfo{volume}{10}}, \bibinfo{pages}{6} (\bibinfo{year}{2019}).

\bibitem{leroy2006percutaneous}
\bibinfo{author}{Leroy, A.} \emph{et~al.}
\newblock \bibinfo{journal}{\bibinfo{title}{Percutaneous renal puncture,
  requirements and preliminary results}}.
\newblock {\emph{\JournalTitle{arXiv preprint physics/0610209}}}
  (\bibinfo{year}{2006}).

\bibitem{fu2021biomechanically}
\bibinfo{author}{Fu, Y.} \emph{et~al.}
\newblock \bibinfo{journal}{\bibinfo{title}{Biomechanically constrained non
  rigid {MR-TRUS} prostate registration using deep learning based 3d point
  cloud matching}}.
\newblock {\emph{\JournalTitle{Medical image analysis}}}
  \textbf{\bibinfo{volume}{67}}, \bibinfo{pages}{101845}
  (\bibinfo{year}{2021}).

\bibitem{icp1987}
\bibinfo{author}{Arun, K.~S.}, \bibinfo{author}{Huang, T.~S.} \&
  \bibinfo{author}{Blostein, S.~D.}
\newblock \bibinfo{journal}{\bibinfo{title}{Least-squares fitting of two 3-d
  point sets}}.
\newblock {\emph{\JournalTitle{IEEE Transactions on Pattern Analysis and
  Machine Intelligence}}} \textbf{\bibinfo{volume}{PAMI-9}},
  \bibinfo{pages}{698--700}, \url{https://doi.org/10.1109/TPAMI.1987.4767965}
  (\bibinfo{year}{1987}).

\bibitem{Agamennoni2016}
\bibinfo{author}{Agamennoni, G.}, \bibinfo{author}{Fontana, S.},
  \bibinfo{author}{Siegwart, R.~Y.} \& \bibinfo{author}{Sorrenti, D.~G.}
\newblock \bibinfo{title}{Point clouds registration with probabilistic data
  association}.
\newblock In \emph{\bibinfo{booktitle}{2016 IEEE/RSJ International Conference
  on Intelligent Robots and Systems (IROS)}}, \bibinfo{pages}{4092--4098},
  \url{https://doi.org/10.1109/IROS.2016.7759602} (\bibinfo{year}{2016}).

\bibitem{hirose2021}
\bibinfo{author}{Hirose, O.}
\newblock \bibinfo{journal}{\bibinfo{title}{A bayesian formulation of coherent
  point drift}}.
\newblock {\emph{\JournalTitle{IEEE Transactions on Pattern Analysis and
  Machine Intelligence}}} \textbf{\bibinfo{volume}{43}},
  \bibinfo{pages}{2269--2286}, \url{https://doi.org/10.1109/TPAMI.2020.2971687}
  (\bibinfo{year}{2021}).

\bibitem{yaoki2019}
\bibinfo{author}{Aoki, Y.}, \bibinfo{author}{Goforth, H.},
  \bibinfo{author}{Arun~Srivatsan, R.} \& \bibinfo{author}{Lucey, S.}
\newblock \bibinfo{title}{Pointnetlk: Robust and efficient point cloud
  registration using pointnet}.
\newblock In \emph{\bibinfo{booktitle}{The IEEE Conference on Computer Vision
  and Pattern Recognition (CVPR)}} (\bibinfo{year}{2019}).

\bibitem{Sarode2019}
\bibinfo{author}{Sarode, V.} \emph{et~al.}
\newblock \bibinfo{title}{Pcrnet: Point cloud registration network using
  pointnet encoding} (\bibinfo{year}{2019}).

\bibitem{Baig2012}
\bibinfo{author}{Baig, A.}, \bibinfo{author}{Chaudhry, M.~A.} \&
  \bibinfo{author}{Mahmood, A.}
\newblock \bibinfo{title}{Local normalized cross correlation for
  geo-registration}.
\newblock In \emph{\bibinfo{booktitle}{Proceedings of 2012 9th International
  Bhurban Conference on Applied Sciences and Technology (IBCAST)}},
  \bibinfo{pages}{70--74}, \url{https://doi.org/10.1109/IBCAST.2012.6177529}
  (\bibinfo{year}{2012}).

\bibitem{den1993calculation}
\bibinfo{author}{den Brinker, A.~C.}
\newblock \bibinfo{journal}{\bibinfo{title}{Calculation of the local
  cross-correlation function on the basis of the laguerre transform}}.
\newblock {\emph{\JournalTitle{IEEE transactions on signal processing}}}
  \textbf{\bibinfo{volume}{41}}, \bibinfo{pages}{1980--1982}
  (\bibinfo{year}{1993}).

\bibitem{hale2006fast}
\bibinfo{author}{Hale, D.}
\newblock \bibinfo{title}{Fast local cross-correlations of images}.
\newblock In \emph{\bibinfo{booktitle}{2006 SEG Annual Meeting}}
  (\bibinfo{organization}{OnePetro}, \bibinfo{year}{2006}).

\bibitem{haskins2019learning}
\bibinfo{author}{Haskins, G.} \emph{et~al.}
\newblock \bibinfo{journal}{\bibinfo{title}{Learning deep similarity metric for
  3d mr--trus image registration}}.
\newblock {\emph{\JournalTitle{International journal of computer assisted
  radiology and surgery}}} \textbf{\bibinfo{volume}{14}},
  \bibinfo{pages}{417--425} (\bibinfo{year}{2019}).

\bibitem{Yang2017}
\bibinfo{author}{Yang, X.}, \bibinfo{author}{Kwitt, R.},
  \bibinfo{author}{Styner, M.} \& \bibinfo{author}{Niethammer, M.}
\newblock \bibinfo{journal}{\bibinfo{title}{Quicksilver: Fast predictive image
  registration - a deep learning approach}}.
\newblock {\emph{\JournalTitle{NeuroImage}}} \textbf{\bibinfo{volume}{158}},
  \bibinfo{pages}{378--396},
  \url{https://doi.org/10.1016/j.neuroimage.2017.07.008}
  (\bibinfo{year}{2017}).

\bibitem{Hu2018}
\bibinfo{author}{Hu, Y.} \emph{et~al.}
\newblock \bibinfo{title}{Label-driven weakly-supervised learning for
  multimodal deformable image registration}.
\newblock In \emph{\bibinfo{booktitle}{2018 IEEE 15th International Symposium
  on Biomedical Imaging (ISBI 2018)}}, \bibinfo{pages}{1070--1074},
  \url{https://doi.org/10.1109/ISBI.2018.8363756} (\bibinfo{year}{2018}).

\bibitem{hu2018weakly}
\bibinfo{author}{Hu, Y.} \emph{et~al.}
\newblock \bibinfo{journal}{\bibinfo{title}{Weakly-supervised convolutional
  neural networks for multimodal image registration}}.
\newblock {\emph{\JournalTitle{Medical image analysis}}}
  \textbf{\bibinfo{volume}{49}}, \bibinfo{pages}{1--13} (\bibinfo{year}{2018}).

\bibitem{voxelmorph2019}
\bibinfo{author}{Balakrishnan, G.}, \bibinfo{author}{Zhao, A.},
  \bibinfo{author}{Sabuncu, M.~R.}, \bibinfo{author}{Guttag, J.} \&
  \bibinfo{author}{Dalca, A.~V.}
\newblock \bibinfo{journal}{\bibinfo{title}{Voxelmorph: A learning framework
  for deformable medical image registration}}.
\newblock {\emph{\JournalTitle{IEEE Transactions on Medical Imaging}}}
  \textbf{\bibinfo{volume}{38}}, \bibinfo{pages}{1788--1800},
  \url{https://doi.org/10.1109/TMI.2019.2897538} (\bibinfo{year}{2019}).

\bibitem{CHEN2022102615}
\bibinfo{author}{Chen, J.} \emph{et~al.}
\newblock \bibinfo{journal}{\bibinfo{title}{Transmorph: Transformer for
  unsupervised medical image registration}}.
\newblock {\emph{\JournalTitle{Medical Image Analysis}}}
  \textbf{\bibinfo{volume}{82}}, \bibinfo{pages}{102615},
  \url{https://doi.org/10.1016/j.media.2022.102615} (\bibinfo{year}{2022}).

\bibitem{chen2021mr}
\bibinfo{author}{Chen, Y.} \emph{et~al.}
\newblock \bibinfo{journal}{\bibinfo{title}{Mr to ultrasound image registration
  with segmentation-based learning for hdr prostate brachytherapy}}.
\newblock {\emph{\JournalTitle{Medical Physics}}}
  \textbf{\bibinfo{volume}{48}}, \bibinfo{pages}{3074--3083}
  (\bibinfo{year}{2021}).

\bibitem{karnik2010assessment}
\bibinfo{author}{Karnik, V.~V.} \emph{et~al.}
\newblock \bibinfo{journal}{\bibinfo{title}{Assessment of image registration
  accuracy in three-dimensional transrectal ultrasound guided prostate
  biopsy}}.
\newblock {\emph{\JournalTitle{Medical physics}}}
  \textbf{\bibinfo{volume}{37}}, \bibinfo{pages}{802--813}
  (\bibinfo{year}{2010}).

\bibitem{shruthi2015detection}
\bibinfo{author}{Shruthi, B.}, \bibinfo{author}{Renukalatha, S.} \&
  \bibinfo{author}{Siddappa, M.}
\newblock \bibinfo{journal}{\bibinfo{title}{Detection of kidney abnormalities
  in noisy ultrasound images}}.
\newblock {\emph{\JournalTitle{International Journal of Computer
  Applications}}} \textbf{\bibinfo{volume}{120}} (\bibinfo{year}{2015}).

\bibitem{kettenbach2005robot}
\bibinfo{author}{Kettenbach, J.} \emph{et~al.}
\newblock \bibinfo{journal}{\bibinfo{title}{Robot-assisted biopsy using
  ultrasound guidance: initial results from in vitro tests}}.
\newblock {\emph{\JournalTitle{European radiology}}}
  \textbf{\bibinfo{volume}{15}}, \bibinfo{pages}{765--771}
  (\bibinfo{year}{2005}).

\bibitem{Isensee2020}
\bibinfo{author}{Isensee, F.}, \bibinfo{author}{Jaeger, P.~F.},
  \bibinfo{author}{Kohl, S. A.~A.}, \bibinfo{author}{Petersen, J.} \&
  \bibinfo{author}{Maier-Hein, K.}
\newblock \bibinfo{journal}{\bibinfo{title}{nnu-net: a self-configuring method
  for deep learning-based biomedical image segmentation}}.
\newblock {\emph{\JournalTitle{Nature Methods}}} \textbf{\bibinfo{volume}{18}},
  \bibinfo{pages}{203 -- 211} (\bibinfo{year}{2020}).

\bibitem{chen2021}
\bibinfo{author}{Chen, J.} \emph{et~al.}
\newblock \bibinfo{journal}{\bibinfo{title}{Transunet: Transformers make strong
  encoders for medical image segmentation}}.
\newblock {\emph{\JournalTitle{arXiv preprint arXiv:2102.04306}}}
  (\bibinfo{year}{2021}).

\bibitem{cao2023}
\bibinfo{author}{Cao, H.} \emph{et~al.}
\newblock \bibinfo{title}{Swin-unet: Unet-like pure transformer for medical
  image segmentation}.
\newblock In \emph{\bibinfo{booktitle}{Computer Vision--ECCV 2022 Workshops:
  Tel Aviv, Israel, October 23--27, 2022, Proceedings, Part III}},
  \bibinfo{pages}{205--218} (\bibinfo{organization}{Springer},
  \bibinfo{year}{2023}).

\bibitem{zhou2021}
\bibinfo{author}{Zhou, H.-Y.} \emph{et~al.}
\newblock \bibinfo{journal}{\bibinfo{title}{nnformer: Interleaved transformer
  for volumetric segmentation}}.
\newblock {\emph{\JournalTitle{arXiv preprint arXiv:2109.03201}}}
  (\bibinfo{year}{2021}).

\bibitem{cotr2021}
\bibinfo{author}{Shen, Z.}, \bibinfo{author}{Lin, C.} \&
  \bibinfo{author}{Zheng, S.}
\newblock \bibinfo{journal}{\bibinfo{title}{Cotr: Convolution in transformer
  network for end to end polyp detection}}.
\newblock {\emph{\JournalTitle{2021 7th International Conference on Computer
  and Communications (ICCC)}}} \bibinfo{pages}{1757--1761}
  (\bibinfo{year}{2021}).

\bibitem{themyr2023}
\bibinfo{author}{Themyr, L.}, \bibinfo{author}{Rambour, C.},
  \bibinfo{author}{Thome, N.}, \bibinfo{author}{Collins, T.} \&
  \bibinfo{author}{Hostettler, A.}
\newblock \bibinfo{title}{Full contextual attention for multi-resolution
  transformers in semantic segmentation}.
\newblock In \emph{\bibinfo{booktitle}{2023 IEEE/CVF Winter Conference on
  Applications of Computer Vision (WACV)}}, \bibinfo{pages}{3223--3232},
  \url{https://doi.org/10.1109/WACV56688.2023.00324} (\bibinfo{year}{2023}).

\bibitem{Yu2019}
\bibinfo{author}{Yu, W.} \emph{et~al.}
\newblock \bibinfo{title}{Liver vessels segmentation based on 3d residual
  u-net}.
\newblock In \emph{\bibinfo{booktitle}{2019 IEEE International Conference on
  Image Processing (ICIP)}}, \bibinfo{pages}{250--254},
  \url{https://doi.org/10.1109/ICIP.2019.8802951} (\bibinfo{year}{2019}).

\bibitem{Gibson2018}
\bibinfo{author}{Gibson, E.} \emph{et~al.}
\newblock \bibinfo{journal}{\bibinfo{title}{Automatic multi-organ segmentation
  on abdominal ct with dense v-networks}}.
\newblock {\emph{\JournalTitle{IEEE Transactions on Medical Imaging}}}
  \textbf{\bibinfo{volume}{37}}, \bibinfo{pages}{1822--1834},
  \url{https://doi.org/10.1109/TMI.2018.2806309} (\bibinfo{year}{2018}).

\bibitem{zhou2018}
\bibinfo{author}{Zhou, Z.}, \bibinfo{author}{Rahman~Siddiquee, M.~M.},
  \bibinfo{author}{Tajbakhsh, N.} \& \bibinfo{author}{Liang, J.}
\newblock \bibinfo{title}{Unet++: A nested u-net architecture for medical image
  segmentation}.
\newblock In \emph{\bibinfo{booktitle}{Deep Learning in Medical Image Analysis
  and Multimodal Learning for Clinical Decision Support: 4th International
  Workshop, DLMIA 2018, and 8th International Workshop, ML-CDS 2018, Held in
  Conjunction with MICCAI 2018, Granada, Spain, September 20, 2018, Proceedings
  4}}, \bibinfo{pages}{3--11} (\bibinfo{organization}{Springer},
  \bibinfo{year}{2018}).

\bibitem{boussaid2021shape}
\bibinfo{author}{Boussaid, H.} \& \bibinfo{author}{Rouet, L.}
\newblock \bibinfo{title}{Shape feature loss for kidney segmentation in 3d
  ultrasound images} (\bibinfo{organization}{BMVC}, \bibinfo{year}{2021}).

\bibitem{zeng2020label}
\bibinfo{author}{Zeng, Q.} \emph{et~al.}
\newblock \bibinfo{journal}{\bibinfo{title}{Label-driven magnetic resonance
  imaging (mri)-transrectal ultrasound (trus) registration using weakly
  supervised learning for mri-guided prostate radiotherapy}}.
\newblock {\emph{\JournalTitle{Physics in Medicine \& Biology}}}
  \textbf{\bibinfo{volume}{65}}, \bibinfo{pages}{135002}
  (\bibinfo{year}{2020}).

\bibitem{xu2022polar}
\bibinfo{author}{Xu, X.} \emph{et~al.}
\newblock \bibinfo{journal}{\bibinfo{title}{Polar transform network for
  prostate ultrasound segmentation with uncertainty estimation}}.
\newblock {\emph{\JournalTitle{Medical Image Analysis}}}
  \textbf{\bibinfo{volume}{78}}, \bibinfo{pages}{102418}
  (\bibinfo{year}{2022}).

\bibitem{Cicek2016}
\bibinfo{author}{{\c{C}}i{\c{c}}ek, {\"O}.}, \bibinfo{author}{Abdulkadir, A.},
  \bibinfo{author}{Lienkamp, S.~S.}, \bibinfo{author}{Brox, T.} \&
  \bibinfo{author}{Ronneberger, O.}
\newblock \bibinfo{title}{3d u-net: Learning dense volumetric segmentation from
  sparse annotation}.
\newblock In \bibinfo{editor}{Ourselin, S.}, \bibinfo{editor}{Joskowicz, L.},
  \bibinfo{editor}{Sabuncu, M.~R.}, \bibinfo{editor}{Unal, G.} \&
  \bibinfo{editor}{Wells, W.} (eds.) \emph{\bibinfo{booktitle}{Medical Image
  Computing and Computer-Assisted Intervention -- MICCAI 2016}},
  \bibinfo{pages}{424--432} (\bibinfo{publisher}{Springer International
  Publishing}, \bibinfo{address}{Cham}, \bibinfo{year}{2016}).

\bibitem{Milletari2016}
\bibinfo{author}{Milletar{\`i}, F.}, \bibinfo{author}{Navab, N.} \&
  \bibinfo{author}{Ahmadi, S.-A.}
\newblock \bibinfo{journal}{\bibinfo{title}{V-net: Fully convolutional neural
  networks for volumetric medical image segmentation}}.
\newblock {\emph{\JournalTitle{2016 Fourth International Conference on 3D
  Vision (3DV)}}} \bibinfo{pages}{565--571} (\bibinfo{year}{2016}).

\bibitem{horstmann2022orientation}
\bibinfo{author}{Horstmann, T.}, \bibinfo{author}{Zettinig, O.},
  \bibinfo{author}{Wein, W.} \& \bibinfo{author}{Prevost, R.}
\newblock \bibinfo{title}{Orientation estimation of abdominal ultrasound images
  with multi-hypotheses networks}.
\newblock In \emph{\bibinfo{booktitle}{Medical Imaging with Deep Learning}}
  (\bibinfo{year}{2022}).

\bibitem{markova2022global}
\bibinfo{author}{Markova, V.}, \bibinfo{author}{Ronchetti, M.},
  \bibinfo{author}{Wein, W.}, \bibinfo{author}{Zettinig, O.} \&
  \bibinfo{author}{Prevost, R.}
\newblock \bibinfo{title}{Global multi-modal 2d/3d registration via local
  descriptors learning}.
\newblock In \emph{\bibinfo{booktitle}{Medical Image Computing and Computer
  Assisted Intervention--MICCAI 2022: 25th International Conference, Singapore,
  September 18--22, 2022, Proceedings, Part VI}}, \bibinfo{pages}{269--279}
  (\bibinfo{organization}{Springer}, \bibinfo{year}{2022}).

\bibitem{Baum2023}
\bibinfo{author}{Baum, Z. M.~C.}, \bibinfo{author}{Saeed, S.~U.},
  \bibinfo{author}{Min, Z.}, \bibinfo{author}{Hu, Y.} \&
  \bibinfo{author}{Barratt, D.~C.}
\newblock \bibinfo{title}{{MR to Ultrasound Registration for Prostate Challenge
  - Dataset}}, \url{https://doi.org/10.5281/zenodo.7870105}
  (\bibinfo{year}{2023}).

\bibitem{Fedorov2015}
\bibinfo{author}{Andrey, F.} \emph{et~al.}
\newblock \bibinfo{journal}{\bibinfo{title}{Open-source image registration for
  {MRI-TRUS} fusion-guided prostate interventions}}.
\newblock {\emph{\JournalTitle{International Journal of Computer Assisted
  Radiology and Surgery}}} \textbf{\bibinfo{volume}{10}},
  \bibinfo{pages}{925--934} (\bibinfo{year}{2015}).

\bibitem{Fedorov2012}
\bibinfo{author}{Andrey, F.} \emph{et~al.}
\newblock \bibinfo{journal}{\bibinfo{title}{3d slicer as an image computing
  platform for the quantitative imaging network}}.
\newblock {\emph{\JournalTitle{Magnetic Resonance Imaging}}}
  \textbf{\bibinfo{volume}{30}}, \bibinfo{pages}{1323--1341},
  \url{https://doi.org/10.1016/j.mri.2012.05.001} (\bibinfo{year}{2012}).
\newblock \bibinfo{note}{Funding Information: We would like to thank all
  current and past users and developers of 3D Slicer for their contribution to
  this software. The authors have been supported in part by the following NIH
  grants. BWH: U01CA151261, P41EB015898, P41RR13218, U54EB005149 and
  1R01CA111288 ; University of Iowa: U01-CA140206 ; GE: P41RR13218 and
  U54EB005149 ; MGH: 1U01CA154601-01 and 4R00LM009889-03 . We are grateful to
  the various agencies and programs that funded support and development of 3D
  Slicer over the years.}

\bibitem{yang2018automatic}
\bibinfo{author}{Yang, G.} \emph{et~al.}
\newblock \bibinfo{title}{Automatic segmentation of kidney and renal tumor in
  ct images based on 3d fully convolutional neural network with pyramid pooling
  module}.
\newblock In \emph{\bibinfo{booktitle}{2018 24th International Conference on
  Pattern Recognition (ICPR)}}, \bibinfo{pages}{3790--3795}
  (\bibinfo{organization}{IEEE}, \bibinfo{year}{2018}).

\bibitem{daniel2021automated}
\bibinfo{author}{Daniel, C.~E., Alexander J.and~Buchanan} \emph{et~al.}
\newblock \bibinfo{journal}{\bibinfo{title}{Automated renal segmentation in
  healthy and chronic kidney disease subjects using a convolutional neural
  network}}.
\newblock {\emph{\JournalTitle{Magnetic Resonance in Medicine}}}
  \textbf{\bibinfo{volume}{86}}, \bibinfo{pages}{1125--1136}
  (\bibinfo{year}{2021}).

\bibitem{staple2004}
\bibinfo{author}{Warfield, S.}, \bibinfo{author}{Zou, K.~H.} \&
  \bibinfo{author}{Wells, W.~M.}
\newblock \bibinfo{journal}{\bibinfo{title}{Simultaneous truth and performance
  level estimation (staple): an algorithm for the validation of image
  segmentation}}.
\newblock {\emph{\JournalTitle{IEEE Transactions on Medical Imaging}}}
  \textbf{\bibinfo{volume}{23}}, \bibinfo{pages}{903--921}
  (\bibinfo{year}{2004}).

\bibitem{Ong2009}
\bibinfo{author}{Ong, R.~E.}, \bibinfo{author}{Glisson, C.~L.},
  \bibinfo{author}{Herrell, S.~D.}, \bibinfo{author}{Miga, M.~I.} \&
  \bibinfo{author}{Galloway, R.}
\newblock \bibinfo{title}{A deformation model for non-rigid registration of the
  kidney}.
\newblock In \emph{\bibinfo{booktitle}{Medical Imaging 2009: Visualization,
  Image-Guided Procedures, and Modeling}}, vol. \bibinfo{volume}{7261},
  \bibinfo{pages}{1022--1030} (\bibinfo{organization}{SPIE},
  \bibinfo{year}{2009}).

\bibitem{wein2008automatic}
\bibinfo{author}{Wein, W.}, \bibinfo{author}{Brunke, S.},
  \bibinfo{author}{Khamene, A.}, \bibinfo{author}{Callstrom, M.~R.} \&
  \bibinfo{author}{Navab, N.}
\newblock \bibinfo{journal}{\bibinfo{title}{Automatic ct-ultrasound
  registration for diagnostic imaging and image-guided intervention}}.
\newblock {\emph{\JournalTitle{Medical image analysis}}}
  \textbf{\bibinfo{volume}{12}}, \bibinfo{pages}{577--585}
  (\bibinfo{year}{2008}).

\bibitem{Dubuisson1994}
\bibinfo{author}{Dubuisson, M.-P.} \& \bibinfo{author}{Jain, A.~K.}
\newblock \bibinfo{title}{A modified hausdorff distance for object matching}.
\newblock In \emph{\bibinfo{booktitle}{Proceedings of 12th International
  Conference on Pattern Recognition}}, vol.~\bibinfo{volume}{1},
  \bibinfo{pages}{566--568 vol.1},
  \url{https://doi.org/10.1109/ICPR.1994.576361} (\bibinfo{year}{1994}).

\bibitem{reinhold2019evaluating}
\bibinfo{author}{Reinhold, J.~C.}, \bibinfo{author}{Dewey, B.~E.},
  \bibinfo{author}{Carass, A.} \& \bibinfo{author}{Prince, J.~L.}
\newblock \bibinfo{title}{Evaluating the impact of intensity normalization on
  mr image synthesis}.
\newblock In \emph{\bibinfo{booktitle}{Medical Imaging 2019: Image
  Processing}}, vol. \bibinfo{volume}{10949}, \bibinfo{pages}{890--898}
  (\bibinfo{organization}{SPIE}, \bibinfo{year}{2019}).

\bibitem{jacobsen2019analysis}
\bibinfo{author}{Jacobsen, N.} \emph{et~al.}
\newblock \bibinfo{journal}{\bibinfo{title}{Analysis of intensity normalization
  for optimal segmentation performance of a fully convolutional neural
  network}}.
\newblock {\emph{\JournalTitle{Zeitschrift f{\"u}r Medizinische Physik}}}
  \textbf{\bibinfo{volume}{29}}, \bibinfo{pages}{128--138}
  (\bibinfo{year}{2019}).

\bibitem{cardoso2022monai}
\bibinfo{author}{Cardoso, M.~J.} \emph{et~al.}
\newblock \bibinfo{title}{Monai: An open-source framework for deep learning in
  healthcare} (\bibinfo{year}{2022}).
\newblock \eprint{2211.02701}.

\bibitem{lewiner2003efficient}
\bibinfo{author}{Lewiner, T.}, \bibinfo{author}{Lopes, H.},
  \bibinfo{author}{Vieira, A.~W.} \& \bibinfo{author}{Tavares, G.}
\newblock \bibinfo{journal}{\bibinfo{title}{Efficient implementation of
  marching cubes' cases with topological guarantees}}.
\newblock {\emph{\JournalTitle{Journal of graphics tools}}}
  \textbf{\bibinfo{volume}{8}}, \bibinfo{pages}{1--15} (\bibinfo{year}{2003}).

\bibitem{du2020robust}
\bibinfo{author}{Du, S.} \emph{et~al.}
\newblock \bibinfo{journal}{\bibinfo{title}{Robust rigid registration algorithm
  based on pointwise correspondence and correntropy}}.
\newblock {\emph{\JournalTitle{Pattern Recognition Letters}}}
  \textbf{\bibinfo{volume}{132}}, \bibinfo{pages}{91--98}
  (\bibinfo{year}{2020}).

\bibitem{figueroa2014biomechanical}
\bibinfo{author}{Figueroa-Garcia, I.}, \bibinfo{author}{Peyrat, J.-M.},
  \bibinfo{author}{Hamarneh, G.} \& \bibinfo{author}{Abugharbieh, R.}
\newblock \bibinfo{title}{Biomechanical kidney model for predicting tumor
  displacement in the presence of external pressure load}.
\newblock In \emph{\bibinfo{booktitle}{2014 IEEE 11th International Symposium
  on Biomedical Imaging (ISBI)}}, \bibinfo{pages}{810--813}
  (\bibinfo{organization}{IEEE}, \bibinfo{year}{2014}).

\bibitem{leroy2007intensity}
\bibinfo{author}{Leroy, A.}, \bibinfo{author}{Mozer, P.},
  \bibinfo{author}{Payan, Y.} \& \bibinfo{author}{Troccaz, J.}
\newblock \bibinfo{journal}{\bibinfo{title}{Intensity-based registration of
  freehand 3d ultrasound and ct-scan images of the kidney}}.
\newblock {\emph{\JournalTitle{International journal of computer assisted
  radiology and surgery}}} \textbf{\bibinfo{volume}{2}},
  \bibinfo{pages}{31--41} (\bibinfo{year}{2007}).

\end{thebibliography}

\section*{Acknowledgements}

This work has received funding from France's region Grand Est. We greatly thank the IHU Strasbourg for its contribution to organizing and implementing the clinical protocol for acquiring the TRUSTED dataset. We thank the radiology department of the NHC Strasbourg for its contribution to the acquisition. We thank Siemens for lending of equipment. We greatly thank the IRCAD France and Africa annotation team (Yvonne Keeza, Grace Ufitinema, and Florien Ujemurwego) for their invaluable participation and work on this study.

\section*{Author contributions statement}
Conceptualization: W.N, C.F., N.T., J.M, A.H, T.C
Methodology: W.N., L.T., N.T., D.G., A.H., T.C.
Software: W.N., L.T., T.C.
Validation: W.N., L.T., T.C.
Formal analysis: W.N.
Investigation: W.N., C.F., L.T.
Resources: P-T. P., A. M., J.M., A.H.
Data curation: W.N., C.F., Y.K.
Writing - Original Draft: W.N., T.C.
Writing - Review \& Editing: N.T., D.G., A.H., T.C.
Visualization: W.N.
Supervision: N.T., D.G., T.C.
Project administration: N.T., D.G., A.H., T.C.
Funding acquisition: J.M., A.H

\section*{Competing interests}

%\textcolor{red}{The corresponding author is responsible for providing a} \href{https://www.nature.com/sdata/policies/editorial-and-publishing-policies#competing}{competing interests statement} \textcolor{red}{on behalf of all authors of the paper. This statement must be included in the submitted article file.}

The authors declare no competing interests.

\section*{Ethical Approval}

The US and CT image data was collected in the clinical study with ID RCB: 2020-A01029-30/SI:20.05.01.539110, which was ethically approved by the French government's "Comité de Protection des Personnes Sud - Mediterranée II".

\end{document}